%% file: main_manuscript.tex
\documentclass[11pt, a4paper]{article}
\usepackage[ margin=1in]{geometry}
\usepackage[affil-it]{authblk}
\usepackage{setspace}
\usepackage[sort&compress,numbers]{natbib}
\usepackage[english]{babel}
\usepackage{graphicx}
\usepackage{mathptmx}
\usepackage{verbatim}
\usepackage{graphicx}
\usepackage{amsmath}
\usepackage{amssymb}
\usepackage{amsfonts}
\usepackage{mathrsfs}
\usepackage{amsmath} % needed for subequations  
\usepackage{romannum}
\usepackage{color}
\usepackage{graphicx}
\usepackage{bm} % bold maths\Delta_0
\usepackage{amssymb}
\usepackage{xspace}
\usepackage{color}
\usepackage{float}
\usepackage{bbold}
\usepackage{tabu}
\usepackage{textcomp}
\usepackage{dcolumn}% Align table columns on decimal point

\newcommand{\angstrom}{\mbox{\normalfont\AA}}
\newcolumntype{L}[1]{>{\raggedright\arraybackslash}m{#1}}
\newcolumntype{C}[1]{>{\centering\arraybackslash}m{#1}}
\newcolumntype{R}[1]{>{\raggedleft\arraybackslash}m{#1}}
\newcolumntype{N}{@{}m{0pt}@{}}

\usepackage{graphicx}% Include figure files
\usepackage{dcolumn}% Align table columns on decimal point
\usepackage{bm}% bold math
\usepackage{epstopdf}
\usepackage{xr}
\usepackage{amsmath}
\usepackage{color}
\usepackage{xcolor}
\usepackage[normalem]{ulem}
\usepackage{wasysym}
\usepackage{siunitx}
\usepackage{physics}
\usepackage{lineno}
\usepackage{gensymb}
\usepackage[T1]{fontenc}
\usepackage{booktabs}
\usepackage[section]{placeins}
\usepackage{textcomp}

\usepackage{caption}
\usepackage[colorlinks=true, linkcolor=black, urlcolor=blue, citecolor=blue]{hyperref}
\usepackage{blindtext} %!
\usepackage[absolute]{textpos}

\makeatletter
\newcommand{\moire}{moir\'e }
\g@addto@macro\@floatboxreset\centering
\makeatother

\begin{document}
\pagenumbering{arabic}
\title{\vspace{-2cm}\normalfont \textbf{Bulk valley transport and Berry curvature spreading at the edge of flat bands}}

\renewcommand\Authfont{\small}

\author[1$\dagger$]{Subhajit Sinha}
\author[1$\dagger$*]{Pratap Chandra Adak}
\author[1]{R.S. Surya Kanthi}
\affil[1]{Department of Condensed Matter Physics and Materials Science, Tata Institute of Fundamental Research, Mumbai 400005, India}
\author[2]{Bheema Lingam Chittari}
\affil[2]{Department of Physics, University of Seoul, Seoul 02504, Korea}
\author[1]{L. D. Varma Sangani}
\author[3]{Kenji Watanabe}
\affil[3]{National Institute for Materials Science, 1-1 Namiki, Tsukuba 305-0044, Japan}
\author[3]{Takashi Taniguchi}
\author[2]{Jeil Jung}
\author[1*]{Mandar M. Deshmukh}
\affil[$\dagger$]{\textnormal{These two authors contributed equally}}
\affil[*]{\textnormal{deshmukh@tifr.res.in (MMD), pratapchandraadak@gmail.com (PCA)}}

\date{}

\maketitle

\begin{abstract}
2D materials based superlattices have emerged as a promising platform to modulate band structure and its symmetries. In particular, \moire periodicity in twisted graphene systems produces flat Chern bands. The recent observation of anomalous Hall effect (AHE) and orbital magnetism in twisted bilayer graphene has been associated with spontaneous symmetry breaking of such Chern bands. However, the valley Hall state as a precursor of AHE state, when time-reversal symmetry is still protected, has not been observed. Our work probes this precursor state using the valley Hall effect. We show that broken inversion symmetry in twisted double bilayer graphene (TDBG) facilitates the generation of bulk valley current by reporting the first experimental evidence of nonlocal transport in a nearly flat band system. Despite the spread of Berry curvature hotspots and reduced quasiparticle velocities of the carriers in these flat bands, we observe large nonlocal voltage several micrometers away from the charge current path -- this persists when the Fermi energy lies inside a gap with large Berry curvature. The high sensitivity of the nonlocal voltage to gate tunable carrier density and gap modulating perpendicular electric field makes TDBG an attractive platform for valley-twistronics based on flat bands.
\end{abstract}
\clearpage

The advancement in twistronics
%~\cite{kim_van_2016,cao_superlattice-induced_2016} 
has opened up new avenues to study electron correlations physics such as Mott insulator states~\cite{kim_tunable_2017-2,cao_correlated_2018,cao_unconventional_2018}, superconductivity~\cite{cao_unconventional_2018,lu_superconductors_2019} and orbital ferromagnetism~\cite{sharpe_emergent_2019-1,serlin_intrinsic_2020} in twisted bilayer graphene (TBG). Recent experiments in twisted double bilayer graphene (TDBG)~\cite{burg_correlated_2019, shen_observation_2019_nature, adak_tunable_2020_prb, liu_spin-polarized_2019,cao_electric_2019} and trilayer graphene aligned to hexagonal boron nitride~(hBN)~\cite{chen_evidence_2019,chen_signatures_2019,chen_tunable_2020} also reveal correlation effects.
While  low energy flat bands enhance electronic correlations ~\cite{suarez_morell_flat_2010, lopes_dos_santos_continuum_2012,  bistritzer_moire_2011}, such \moire systems support topological bands with nonzero Chern number~\cite{song_topological_2015, sharpe_emergent_2019-1, serlin_intrinsic_2020, zhang_nearly_2019, chen_tunable_2020}.
In fact, the observation of anomalous Hall state in TBG~\cite{sharpe_emergent_2019-1,serlin_intrinsic_2020} has been explained by spontaneous symmetry breaking of degenerate Chern bands. 
Such observations point to rich topology in twisted systems  governed by nonzero Berry curvature and invite probing of Berry curvature induced physics~\cite{xiao_berry_2010-2,nagaosa_anomalous_2010-1}. 
We present direct experimental evidence of bulk valley transport due to Berry curvature hotspots in flat bands, an aspect that has been little explored.

%\textbf{Explaining VHE}
%.  (VHE)~\cite{JSong_bulk_valley_current,gorbachev_detecting_2014,shimazaki_generation_2015-1,sui_gate-tunable_2015-1,song_topological_2015}   
When inversion symmetry is broken, two-dimensional honeycomb lattices with time-reversal symmetry can have nonzero Berry curvature of same magnitude but opposite sign in two degenerate valleys, $K$  and $K'$.
The nonzero Berry curvature can manifest itself in bulk valley transport via valley Hall effect (VHE) as electrons from two valleys are deflected to two opposite directions perpendicular to the in-plane electric field~\cite{xiao_berry_2010-2, nagaosa_anomalous_2010-1}. 
In systems such as graphene with small inter-valley scattering, the valley current can be detected by an inverse VHE at probes away from the charge current path in the form of a nonlocal resistance~\cite{gorbachev_detecting_2014, shimazaki_generation_2015-1, sui_gate-tunable_2015-1}.
Pure bulk valley current has been generated and detected in \moire system of monolayer graphene aligned to hBN~\cite{gorbachev_detecting_2014}. 
%Here the lattice mismatch breaks the inversion symmetry and creates large Berry curvature hotspots near the charge neutrality point (CNP) and  the secondary Dirac points (SDP).
Similar nonlocal response has been observed in insulating systems like gapped bilayer graphene~\cite{shimazaki_generation_2015-1, sui_gate-tunable_2015-1}  with the insensitivity to device edge details suggesting bulk transport. 
%In bilayer graphene, the perpendicular electric field does the job of breaking inversion symmetry while opening up a gap at CNP~\cite{shimazaki_generation_2015-1, sui_gate-tunable_2015-1}. 
In both the systems, nonlocal resistance has been observed near the Berry curvature hotspots.

%\textbf{Motivation behind our work in TDBG and summarizing the work}
In this work, we investigate twisted double bilayer graphene (TDBG) where two copies of Bernal-stacked bilayer graphene are put on top of each other with a small twist angle. 
%The band structure of small-angle TDBG can be tuned by perpendicular electric field~\cite{burg_correlated_2019,adak_tunable_2020_prb}. 
While the electric field tuned  \moire bands in TDBG have finite Berry curvature, the associated Chern number can be nonzero making it an interesting platform for hosting valley current~\cite{chebrolu_flat_2019, zhang_nearly_2019, koshino_band_2019-1, lee_theory_2019, choi_intrinsic_2019, liu_quantum_2019-2}. 
We measure multiple TDBG devices and observe large nonlocal resistance whenever the Fermi energy lies in the gap -- the charge neutrality point~(CNP) gap or the \moire gaps. 
We explore the dependence of the nonlocal resistance on electric field, charge density and temperature in detailed measurements. 
Our analysis finds evidence that the nonlocal resistance originates from bulk valley transport, while at low temperature edge transport starts playing a role. 
%We further point out the imprint of flat band in the measured nonlocal resistance due to the momentum spreading of the Berry curvature. The observation of valley current in twisted system is not only key to understand the underlying band topology, but also such valley current can be utilized in valleytronics. 
Twistronic system, like the one we present, offers two key knobs for bulk valleytronics-- firstly, the magnitude of Berry curvature is inversely related to the gap, and secondly, the tunability of Fermi velocity tunes the sharpness of the Berry curvature hotspot.

\begin{figure}[hbt]
	\begin{center}

	\includegraphics[width=15.5cm]{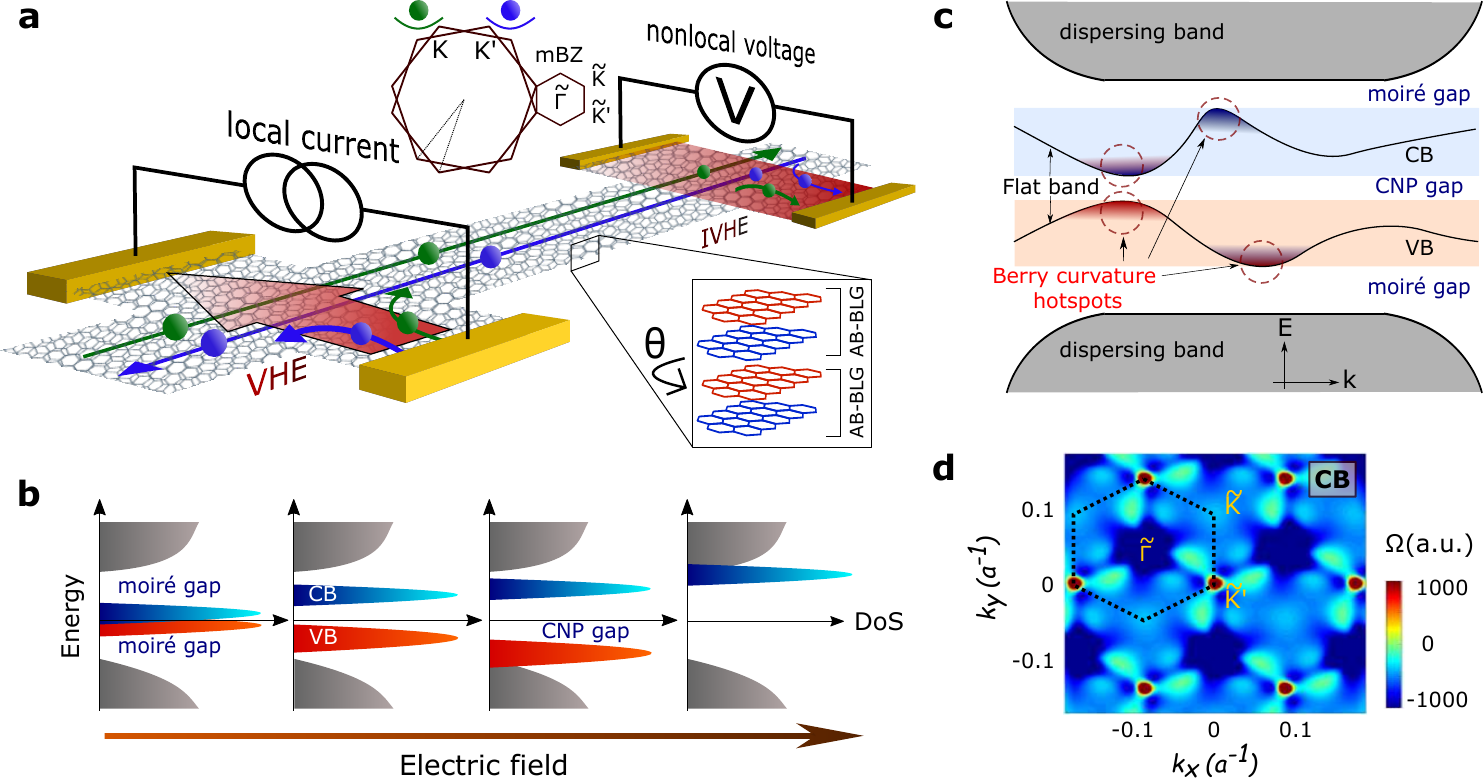}
	
	\caption{ \label{fig:fig1}  \textbf{Nonlocal measurement scheme and electric field tunability in TDBG.} \textbf{a}, Nonlocal measurement scheme of bulk valley current. A net valley current is generated by the charge current through the local probes in its transverse direction due to nonzero Berry curvature via valley Hall effect (VHE). In the nonlocal probes, the valley current generates a voltage drop by inverse VHE. 
			\textbf{b}, A schematic depicting the electric field tunable \moire bands  in twisted double bilayer graphene (TDBG). Two low energy flat bands - the conduction band (CB in blue) and the valence band (VB in red) are separated from high energy dispersing bands (grey) by two electric field tunable \moire gaps. CNP gap opens up between two flat bands as electric field is increased. 
			\textbf{c}, Schematic of a band structure of TDBG at a finite electric field with locations of the Berry curvature hotspots encircled. 
			\textbf{d}, A map of calculated Berry curvature $\Omega$ in $k$-space for $K$ valley of the conduction flat band at a finite electric field. Here $a$ is the lattice constant of graphene. This map shows the hotspots located at the symmetry points $\tilde{\Gamma}$, $\tilde{K}$, and $\tilde{K'}$ in the \moire Brillouin zone. The Berry curvature for $K'$ valley has same magnitude but opposite sign.}
\end{center}
\end{figure}

%\section{Results}
%\textbf{TDBG devices and explaining the nonlocal measurement scheme}
For detecting bulk valley current, we follow a measurement scheme similar to that used for detecting spin current in spintronics devices~\cite{valenzuela_direct_2006, abanin_nonlocal_2009}, as shown in Fig.~\ref{fig:fig1}a.
A finite charge current $I$ is passed using two local probes at two opposite sides of the device channel. 
VHE drives a valley current along the channel and a voltage, $V_{\text{NL}}$ is generated in the nonlocal probes by inverse VHE. 
We quantify this as nonlocal resistance $R_\text{NL} = V_{\text{NL}}/I$.
%As shown in Fig.~\ref{fig:fig1}b,  
We independently control both the charge density $n$ and the perpendicular electric displacement field $D$ aided by the dual-gated structure of our devices using a metal top gate and highly doped silicon back gate (see Methods).
%given by $n=(C_\text{TG}V_\text{TG}+C_\text{BG}V_\text{BG})/e$ and $D=(C_\text{TG}V_\text{TG}-C_\text{BG}V_\text{BG})/2$, where $C_\text{TG}$ and $C_\text{BG}$ are the capacitance per unit area of the top and the back gate respectively, and $e$ is the charge of an electron.

The perpendicular electric field has a profound effect on the band structure of TDBG~\cite{shen_observation_2019_nature, liu_spin-polarized_2019, cao_electric_2019, burg_correlated_2019, adak_tunable_2020_prb}. 
As depicted in Fig.~\ref{fig:fig1}b, at zero electric field, the system has low energy flat bands separated from higher energy dispersing bands by two \moire gaps. 
As the electric field is increased, a gap opens up at the CNP separating two flat bands. 
The \moire gaps close sequentially upon further increase of the electric field.
In Fig.~\ref{fig:fig1}c we present a schematic of the band structure at finite electric field and show the existence of Berry curvature hotspots in the flat bands. 
The color scale plot of calculated Berry curvature in Fig.~\ref{fig:fig1}d depicts the locations of hotspots in the $k$-space of the conduction band for $K$ valley. Details of band structure and Berry curvature map is provided in Supplementary Sec.~\ref{sec:theory}.

\begin{figure}[hbt]
	\begin{center}
	\includegraphics[width=15cm]{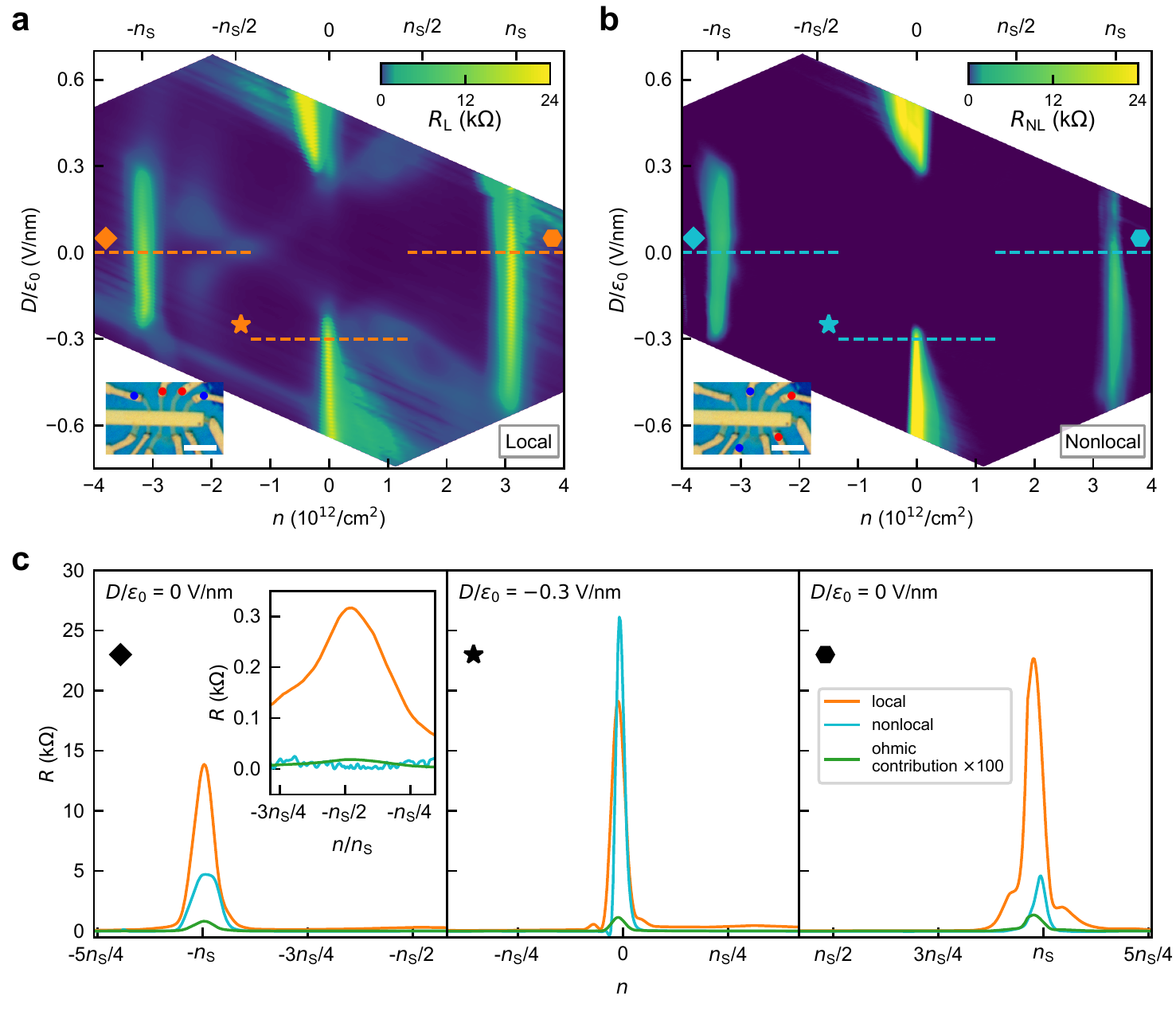}
	\caption{ \label{fig:fig2}  \textbf{Local and nonlocal resistance in TDBG device with twist angle 1.18$\degree$ at 1.5~K.}  \textbf{a, b}, Local and nonlocal resistance as a function of charge density and  electric field. 
			The dashed overlay line shows the location of line slices plotted in \textbf{c}. 
			$n_\text{S}$ denotes the charge density to fill one \moire flat band.  
			Insets show the micrographs of the device with voltage (current) terminals indicated by red (blue) dots. The scale bar corresponds to 5 \textmu m. 
			\textbf{c}, Both local and nonlocal resistance, along with the calculated ohmic contribution. 
			The CNP peak (at $n=0$) is shown for nonzero electric field where the CNP gap is opened. 
			The nonlocal resistance is much larger than the ohmic contribution. %The inset shows the optical microscope image of the device. The scale bar corresponds to 2 \textmu m.
			The inset zooms the resistance variation around $n = -n_\text{S}/2$ at $D/\epsilon_0 = 0$~V/nm, where nonlocal resistance is almost zero in spite of significant local resistance.
	}
\end{center}
\end{figure}

%\textbf{Local vs nonlocal resistance - Fig 2}
We now present the experimental results for a TDBG device with twist angle 1.18$\degree$. 
This device shows a high degree of twist angle homogeneity $\delta \theta \sim 0.05\degree$ over 8 microns; this is crucial for observing nonlocal resistance (Supplementary Sec.~\ref{sec:angle}).
% This device has eight electrical contacts (labeled numerically in the inset of Fig,~\ref{fig:fig2}c). If not mentioned otherwise, in this text we denote $R_{23,14}$ as local resistance and $R_{27,45}$ as nonlocal resistance, where in $R_{ij,kl}$ the indices $i,j$ refer to the voltage probes and the indices $k,l$ refer to the current probes.
In Fig.~\ref{fig:fig2}a, we show a color scale plot of four-probe local resistance as a function of perpendicular electric field and charge density.  
We see large resistance at $n=0$ at high electric field due to gap opening at CNP, and at $n=\pm n_\text{S} = 3.2 \times 10^{12}$~cm$^{-2}$ corresponding to the \moire gaps. 
Here $ n_\text{S}$ is the number of electrons required to fill one flat band. 
In Fig.~\ref{fig:fig2}b, we plot the measured nonlocal resistance which is large only at the gaps.  
Apart from the resistance peak at the gaps, there are other high resistance regions in the local resistance, characteristics to small-angle TDBG~\cite{shen_observation_2019_nature, liu_spin-polarized_2019, cao_electric_2019, burg_correlated_2019, adak_tunable_2020_prb}. Such examples are the cross-like feature originating at $D = 0$ in the hole side and the ring-like regions in the electron side for $|D|/\epsilon_0$ around 0.3 V/nm.  
The absence of these features in the nonlocal signal provides evidence that the nonlocal signal is distinct from the local resistance and is only appreciable when the Fermi energy crosses the gaps that possess large Berry curvature.

In Fig.~\ref{fig:fig2}c,  we plot line slices from the color plots to show both the local and nonlocal resistances as a function of charge density. 
This clearly shows a large nonlocal signal at $n=\pm n_{\text{S}}$, corresponding to the \moire gaps at $D/\epsilon_0 = 0$ (left and right panels). 
In the middle panel, we plot the resistance at CNP for $D/\epsilon_0 =-0.3$ V/nm. 
On the same plot, we additionally plot the ohmic contribution to the nonlocal resistance due to stray current~\cite{abanin_giant_2011}. 
%The ohmic contribution decays exponentially along the length $l$ of the conduction channel governed by the van der Pauw length scale $w/\pi$, $R^{\text{ohm}}_\text{NL} =(\rho_{xx}/\pi)\times \exp(-\pi l/w)$, where $w$ is the width of the conduction channel. 
The calculated ohmic contribution (in Methods), being at least two orders of magnitude lower, cannot account for the large nonlocal resistance we observe.
%Here we note, unlike previous reports of nonlocal resistance in graphene systems, the nonlocal resistance in TDBG does not fall much rapidly compared to the local signal as a function of the charge density. This may be due to the flatness of the underlying bands, a hallmark of small-angle twisted graphene systems.
%We further check that the nonlocal resistance we measure is consistent with reciprocity (see Sec.~V in the SI for details).

%\textbf{Imprint of flat band - broader nonlocal resistance peak}
Now we discuss an interesting difference in nonlocal resistance of TDBG compared to hBN aligned MLG~\cite{gorbachev_detecting_2014} or gapped BLG~\cite{shimazaki_generation_2015-1, sui_gate-tunable_2015-1}. 
In a flat band system, the kinetic energy of the electrons is quenched. As the electrons slow down, with reduced Fermi velocity~$v_\text{F}$, they start to see an enhanced effect of the other energy scales in the system, for example, the $e$-$e$ interaction. 
In a similar way, smaller $v_\text{F}$ renormalizes the gap.  The enhancement of the effective band gap results in the spreading of the Berry curvature hotspots.
To quantitatively understand this effect, we consider the Berry curvature of a gapped (2$\Delta_\text{g}$) MLG with renormalized $v_\text{F}$ to incorporate the effect of band flatness, $|\Omega(k)|=\frac{(\hbar v_\text{F})^2 \Delta_\text{g}}{2[(v_\text{F}\hbar k)^2+\Delta_\text{g}^2]^{3/2}}$. We find that the Berry curvature hotspot extends more in the $k$-space as $v_\text{F}$ is decreased (Supplementary Sec.~\ref{sec:berry}). 
As a result, $R_\text{NL}$ is appreciable over a large range of charge density around the gap in TDBG. 
This is evident in Fig.~\ref{fig:fig2}c as the nonlocal resistance peak is broad in the charge density axis. 
On the other hand, in the earlier reported systems~\cite{gorbachev_detecting_2014, shimazaki_generation_2015-1,sui_gate-tunable_2015-1} nonlocal resistance falls more rapidly than the local resistance, as the charge density is tuned away from the gaps (comparison in Supplementary Sec.~\ref{sec:compare}).

%\textbf{Scaling - Fig 3}
We now proceed to understand the microscopic origin of the nonlocal signal. For diffusive transport of valley polarized electrons through the bulk, the nonlocal resistance $R_\text{NL}$ generated via VHE is given by~\cite{gorbachev_detecting_2014}:
\begin{equation}\label{eqn:1}
	R_{\text{NL}}=\frac{1}{2} \left( \frac{\sigma_{xy}^{\text{VH}}}{\sigma_{xx}}\right)^2 \frac{W}{\sigma_{xx}l_\text{v}} \exp(-\frac{L}{l_\text{v}}).
\end{equation}
Here, $\sigma_{xy}^{\text{VH}}$ is the valley Hall conductivity, $l_\text{v}$ indicates the valley diffusion length, with $L$ and $W$ being the length and the width of the Hall bar channel, respectively. This equation holds good when $\sigma_{xy}^\text{VH}/\sigma_{xx}<<1$, and results in a scaling relation between  the local and the nonlocal resistance, $R_\text{NL} \propto R^3_\text{L}$ with $R_\text{L} = 1/\sigma_{xx}$.

\begin{figure}[hbt]
	\begin{center}
	\includegraphics[width=15cm]{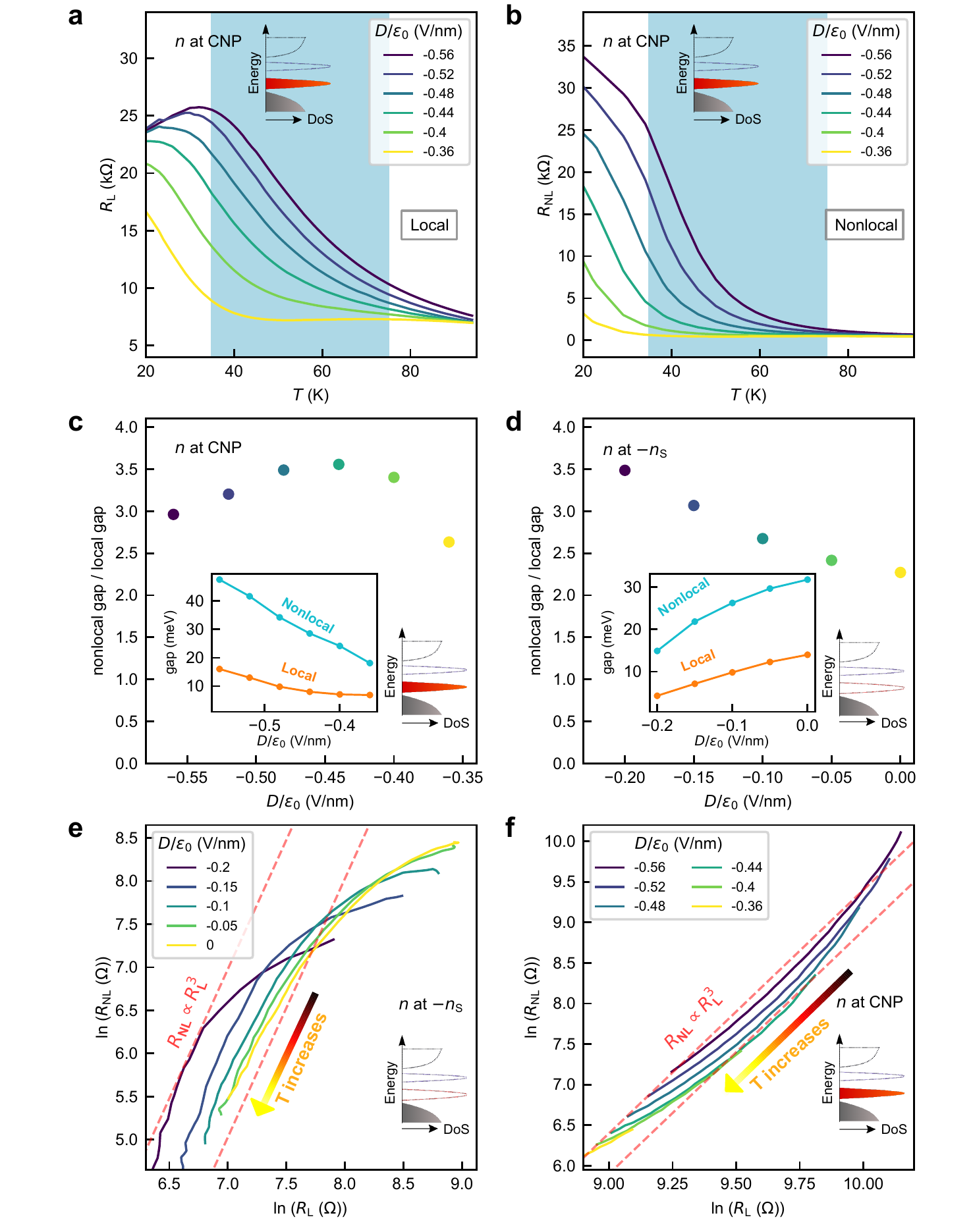}
	\caption{
		\label{fig:fig3}
	 \textbf{Temperature dependence and scaling of local and nonlocal resistance.}  
			Inset of each  panel shows schematic of band filling in that measurement.
			\textbf{a, b}, Variation of local and nonlocal resistance with temperature at CNP for different electric fields. 
			The region shaded with blue color is governed by Arrhenius activation and is the region used to show scaling relation in \textbf{f}. 
			\textbf{c, d}, Ratio of activation gaps for local and nonlocal resistance for CNP~(\textbf{c}) and the \moire peak at $n=-n_\text{S}$~(\textbf{d}). 
			Insets show the variation of gaps with electric field. 
			The ratio, being close to 3, shows that $R_\text{NL} \propto R_\text{L}^3$ and thus strongly supports bulk valley transport.
			\textbf{e, f}, The scaling of nonlocal resistance with local resistance at different electric fields for $n=-n_\text{S}$~(\textbf{e}) and CNP~(\textbf{f}). 
			Here, temperature is used as a parameter to show the scaling.}
\end{center}
\end{figure}

To examine the scaling relation we measure temperature dependence of the local and the nonlocal resistance  for different $D$, as plotted in Fig.~\ref{fig:fig3}a and Fig.~\ref{fig:fig3}b respectively, for the case of CNP. 
The local resistance shows Arrhenius activation behavior due to gaps in the system. 
The nonlocal resistance also follows activation behavior, but with higher gaps than the local resistance. 
In Fig.~\ref{fig:fig3}c and in Fig.~\ref{fig:fig3}d, we plot the ratio of the nonlocal to the local gap as a function of electric field for $n=0$ and $n=-n_\text{S}$, respectively. 
The insets of Fig.~\ref{fig:fig3}c and Fig.~\ref{fig:fig3}d show the values of the activation gaps as a function of electric field. 
Although the individual gaps are tuned by the electric field, the ratio varies within 2.3 to 3.5. The ratio being close to 3 establishes $ R_\text{NL}\propto R_\text{L}^3 $  and hence supports bulk valley transport through equation~(\ref{eqn:1}).
Also, this measurement reinforces our understanding that the contribution of $R_\text{L}$ in $R_\text{NL}$ is minimal.

Now we closely examine the cubic scaling relation as a function of temperature (for scaling as a function of electric field, see Supplementary Sec.~\ref{sec:scaling}). In Fig.~\ref{fig:fig3}e and Fig.~\ref{fig:fig3}f, we plot the nonlocal resistance against the local resistance in logarithmic scale with temperature as a parameter. Fig.~\ref{fig:fig3}e shows the case for $n=-n_\text{S}$, where the temperature varies from 10~K to 75~K. 
The scaling remains cubic, with deviation at low $T$. This low temperature deviation from cubic scaling to being nearly independent of local resistance is consistent with the literature~\cite{shimazaki_generation_2015-1, sui_gate-tunable_2015-1}. At low temperature, the system enters into a large valley Hall angle regime, where the assumption $\sigma_{xy}^\text{VH}/\sigma_{xx}<<1$ is no longer valid~\cite{beconcini_nonlocal_2016}. 

The case where the system is at the charge neutrality is shown in Fig.~\ref{fig:fig3}f (the chosen range of temperature for showing scaling is shaded by blue  in Fig.~\ref{fig:fig3}a and Fig.~\ref{fig:fig3}b). 
The scaling is cubic in intermediate temperature range, consistent with bulk valley transport, with departures at both ends. 
We note that $\sigma_{xy}^{\text{VH}}$ in equation~(\ref{eqn:1}) can have temperature dependence and decrease from its quantized value at elevated temperatures compared to the gap~\cite{shimazaki_generation_2015-1}. 
Such a phenomenon can explain the departure from cubic scaling at the high temperature end. 
The low temperature deviation at CNP is different from that in the case of $-n_\text{S}$, as a transition to higher power laws occurs. We note that the nonlocal signal at CNP originates due to Berry curvature hotspots located at the edge of flat bands (See Supplementary Sec.~\ref{sec:theory} for $\sigma_{xy}^{\text{VH}}$ at CNP in TDBG). At low temperatures, strong $e$-$e$ correlations may give rise to edge states~\cite{jeil_prl_refX1}.

%\section{Conclusion}

Twisted double bilayer graphene offers a unique platform since it provides electrical control over the flatness
of bands through $v_\text{F}$ and the band gap it hosts. 
Our study shows that one can further use this electrical control to induce bulk valley current and offers new opportunities in valleytronics -- via manipulating valley current by the tunable band gaps and the band flatness in twistronics.
In particular, we show that the renormalized velocity in a flat band causes momentum spreading out of the Berry curvature hotspots.  
%We find large nonlocal resistances in the CNP and \moire gaps. 
%Additionally, we demonstrate tuning of this nonlocal signal by the use of perpendicular electric field. 
%We conclude that the nonlocal signal originates due to the bulk transport of valley polarized electrons by establishing cubic scaling relation between nonlocal and local activation gaps and resistances.  
While we observe bulk valley current at elevated temperature, we cannot exclude the possibility that at low temperatures the nonlocal response is additionally mediated by the edge modes associated to the valley Hall effect or other spontaneous quantum Hall effect~\cite{jeil_prl_refX1,jeil_prb_refX2} resulting from flat Chern bands~\cite{zhang_nearly_2019}. 
The recent observation of anomalous Hall effect (AHE) and orbital magnetism in hBN-aligned TBG has been associated to the occupation of an excess valley- and spin-polarized Chern band by spontaneously breaking time-reversal symmetry~\cite{sharpe_emergent_2019-1,serlin_intrinsic_2020}. 
Our demonstration of VHE by nonlocal transport in TDBG, while preserving valley degeneracy, provides strong evidence that VHE state is indeed the parent state of the AHE state. We expect AHE state in TDBG as well, when the valley symmetry is broken.
Additionally, our work opens up new possibilities to explore chargeless valley transport in other \moire systems like trilayer graphene aligned with hBN, twisted trilayer graphene and other twisted transition metal dichalcogenides having topological Chern bands.

\section*{Methods:}
Our dual gated devices are made of hBN/TDBG/hBN stacks on SiO$_2$($\sim$~280~nm)/Si$^{++}$ substrate. To make the stacks, we exfoliate graphene and choose suitable bilayer graphene flakes based on optical contrast and then confirm the layer number by Raman spectroscopy. 
The suitable hBN flakes are selected based on color, and we confirm the thickness by AFM after the stack is completed. Bilayer graphene flakes are sliced into two halves using a tapered optical fiber scalpel, a method reported in~\cite{sangani_facile_2020}. Subsequently, the flakes are assembled using the standard poly(propylene) carbonate (PPC) based dry transfer method~\cite{wang_one-dimensional_2013-1}. The twist angle is introduced by rotating the bottom stage during the pick up of the second half of the graphene.
Subsequently, we define the geometry of the devices by e-beam lithography, followed by  CHF$_3$ + O$_2$ plasma etching. One-dimensional edge contacts to the graphene are made by etching the stack and depositing Cr/Pd/Au. The top gate is made by depositing Cr/Au. 

We fabricate and measure nonlocal transport in multiple devices. The dual-gated structure using a metal top gate and highly doped silicon back gate enables us to have independent control of both the charge density $(n)$ and the perpendicular electric displacement field $(D)$ given by $n=(C_\text{TG}V_\text{TG}+C_\text{BG}V_\text{BG})/e$ and $D=(C_\text{TG}V_\text{TG}-C_\text{BG}V_\text{BG})/2$, where $C_\text{TG}$ and $C_\text{BG}$ are the capacitance per unit area of the top and the back gate respectively, and $e$ is the charge of an electron. All the data reported in the main manuscript are measured using Device 1 with twist angle 1.18$\degree$. 
%This device shows a high degree of twist angle homogeneity $\delta \theta \sim 0.05\degree$ over 8 microns; this is crucial for observing nonlocal resistance (see Sec.~xx in the Supplementary for details). 
We present data from another device with twist angle 1.24$\degree$ in Sec.~\ref{sec:dev2} of the Supplementary. The twist angle $\theta$ is calculated from $n_\text{S}$ where $n_\text{S}=8\theta^{2}/(\sqrt{3}a^{2})$, $a=0.246$~nm is the lattice constant of graphene.

The transport measurements reported in the main text and the supplementary are conducted using a low frequency ($\sim$ 17 Hz) lock-in technique by sending a current $\sim$ 10 nA and measuring the voltage after amplifying using SR560 preamplifier or preamplifier model 1021 by DL instruments, Ithaca. Additionally, we perform dc measurements to verify that the measured nonlocal signal is repeatable and independent of the measurement scheme (Supplementary Sec.~\ref{sec:meas}). We further checked that the nonlocal signal is consistent with reciprocity (Supplementary Sec.~\ref{sec:reciprocity}). 
The ohmic contribution to the nonlocal resistance due to stray current, as plotted in Fig.~2c, has been calculated by using the van der Pauw formula, $R^{\text{ohm}}_\text{NL} =(\rho_{xx}/\pi)\times \exp(-\pi L/W)$~\cite{shimazaki_generation_2015-1}. Here, $L$ and $W$ are the length and width of the conduction channel, respectively. For the device presented in the main text, we choose $L$=4~$\text{\micro}$m and $W$=2~$\text{\micro}$m for performing the nonlocal measurements. This contribution decays exponentially along the length of the conduction channel. For the local measurements, we use the four-probe method and choose both $L$ and $W$ to be 2~$\text{\micro}$m from the same device, allowing us to use the relation $R_\text{L} = 1/\sigma_{xx}$ in the main text.

\input{references.bbl}
%\bibliography{references}
%\printbibliography
%\include{Methods/TDBG-Nonlocal-methods.tex}

\section*{Acknowledgements:}
We thank Allan MacDonald, Justin Song, Rajdeep Sensarma, Vibhor Singh, Sajal Dhara and Biswajit Datta for helpful discussions and comments. We acknowledge Nanomission grant SR/NM/NS-45/2016 and Department of Atomic Energy of Government of India for support. Preparation of hBN single crystals is supported by the Elemental Strategy Initiative conducted by the MEXT, Japan and JSPS KAKENHI Grant Number JP15K21722. This work is supported by the Korean NRF for B.L.C. through Basic Science Research Program of the National Research Foundation of Korea (NRF) funded by the Ministry of Education Grant No. 2018R1A6A1A06024977 and Grant No. NRF-2020R1A2C3009142, and for J.J. through Samsung Science and Technology Foundation under project no. SSTF-BA1802-06.

\section*{Author contributions:}
P.C.A, S.S, R.S.S.K., and L.D.V.S. fabricated the devices. P.C.A. and S.S. did the measurements and analyzed the data. B.L.C. and J.J. did the theoretical calculation. K.W. and T.T. grew the hBN crystals. P.C.A, S.S., and M.M.D. wrote the manuscript with inputs from everyone. M.M.D. supervised the project.

\clearpage

\clearpage	

\clearpage

\clearpage

\begin{center}
\textbf{\huge Supplementary Information}
\end{center}

\renewcommand{\theHsection}{Ssection.\thesection}
\renewcommand{\thesection}{\Roman{section}}
\setcounter{section}{0}

\renewcommand{\theHfigure}{Sfigure.\thefigure}
\renewcommand{\thefigure}{S\arabic{figure}}
\setcounter{figure}{0}

\renewcommand{\theHequation}{Sequation.\theequation}
\renewcommand{\theequation}{S\arabic{equation}}
\setcounter{equation}{0}

\renewcommand{\theHtable}{Stable.\thetable}
\renewcommand{\thetable}{S\Roman{table}}
\setcounter{table}{0}

\section{Calculation of band structure and valley Hall conductivity in TDBG}\label{sec:theory}

As discussed in the theory paper Ref.~\cite{tbbg-raju},  moir\'e bands theory for the moir\'e pattern superlattice \cite{bistritzer2011moire} and the accurate continuum models \cite{jung2014ab} are used to obtain the electronic structure of the twisted double bilayer graphene (TDBG). The continuum model of Bistritzer-MacDonald for the twisted bilayer graphene (TBG)~\cite{bistritzer2011moire} is extended to the case of twisted double bilayer graphene (TDBG), the Hamiltonian of TDBG at the valley K with the interlayer coupling between the twisted layers through a first-harmonic stacking-dependent interlayer tunneling function, and subject to $\Delta_{i}$ intralayer potentials as

\begin{equation}\label{eqn:TDBGmatrix}
H_\textrm{TDBG} (\theta) = \left(\begin{array}{cccc}
h_\textrm{t}^+ + \bar{\Delta}_1 & t_s^+ & 0 & 0  \\
t_s^{+\dagger} & h_\textrm{b}^+ + \bar{\Delta}_2 & T(\boldsymbol{r}) & 0 \\
0 & T^{\dagger}(\boldsymbol{r}) & h_\textrm{t}^- + \bar{\Delta}_3 & t_s^- \\
0 & 0 & t_s^{-\dagger} & h_\textrm{b}^- + \bar{\Delta}_4 \\
\end{array}\right),
\end{equation}

where $h_\textrm{t/b}^{\pm}= h_\textrm{t/b}(\pm\theta/2)$ such that the relative twist angle between the bilayers is $\theta$.

The Dirac Hamiltonian given by $h(\theta)=\upsilon_\text{F} \vec{R}_{-\theta}\boldsymbol{p}\cdot \boldsymbol{\sigma_{xy}}$ includes a phase shift due to a rotation $\vec{R}_{-\theta}$ such that $e^{\pm i \theta_p}\rightarrow e^{\pm i (\theta_p-\theta)}$, where $\sigma_{xy} = (\sigma_x,\sigma_y)$ and $\sigma_z$ are the graphene sublattice pseudospin Pauli matrices, and the momentum is defined in the xy plane $\boldsymbol{p} = (px, py)$, where we assume K valley unless stated otherwise. The Fermi velocity $\upsilon_\text{F} = \upsilon_0$ defined from $\upsilon_i = \sqrt{3}|t_i|a/2\hbar$ is related to the intralayer nearest-neighbor hopping term t$_0$ = -3.1 eV that captures the experimental moir\'e band features better \cite{jung-crommie}. 

The top and bottom bilayer graphene (BG) are labeled through the positive/negative (+/-) rotation signs, while in turn we have top/bottom (t/b) graphene layers within each BG that are coupled through the matrices $t_s^{\pm}$. The interlayer coupling model of a bilayer graphene is given by
\begin{equation}
t^{\pm}_{AB} = \left(\begin{array}{cc}
-\upsilon_4 \pi^{\pm \dagger} & -\upsilon_3 \pi^{\pm}\\
t_1 & -\upsilon_4 \pi^{\pm \dagger} \\
\end{array}\right), \hspace{0.5 cm} t^{\pm}_{BA} = t^{\pm \dagger}_{AB},
\end{equation}

satisfying $t_{s=+}=t^{\dag}_{s=-}$ for AB or BA ($s = \pm 1$) stacking-dependent interlayer coupling that consists of a minimal coupling term $t_s = t_1 (\sigma_x -is\sigma_y )/2$ plus remote hopping contributions through the terms t$_3$ = 0.283, t$_4$ = 0.138 eV, giving rise to trigonal warping and electron-hole asymmetry. The $\pi^{\pm}$ operators include the phases due to $\pm \theta/2$ layer rotation. The Hamiltonian of graphene is given by $h_\textrm{l}^{\pm} (\theta) = h^{}\pm(\theta) + \delta(\mathbb{1}-ls\sigma_z)/2$ where the second term adds a $\delta= $0.015 eV sublattice potential at the higher energy dimer sites at the t/b layers $l = \pm$ \cite{jung-lda-bilayer}, that depends on AB or BA stacking $s = \pm$, respectively.  The site potentials $\bar{\Delta}_{i}$ are mapped on its sublattices through $\bar{\Delta}_{i} = \Delta_{i}\mathbb{1}$  where $i$ = 1, 2, 3, 4 are the layer labels from top to bottom, and $\mathbb{1}$ is a $2 \cross 2$ identity matrix. Here, we have an additional control knob to change the electronic structure through a perpendicular external electric field that modifies the interlayer potential $\Delta_{i}$ values in equation~(\ref{eqn:TDBGmatrix}). The potential drops introduced by an external electric field could be modeled through the parameter set $\Delta_1 = -\Delta_4$, $\Delta_2 = -\Delta_3$, redefined as $\Delta_1 = 3\Delta/2$ and $\Delta_2 = \Delta/2$, where $\Delta$ is the interlayer potential difference between each BG.

\begin{figure*}[hbt]
	\begin{center}
		\includegraphics[width=0.95\textwidth]{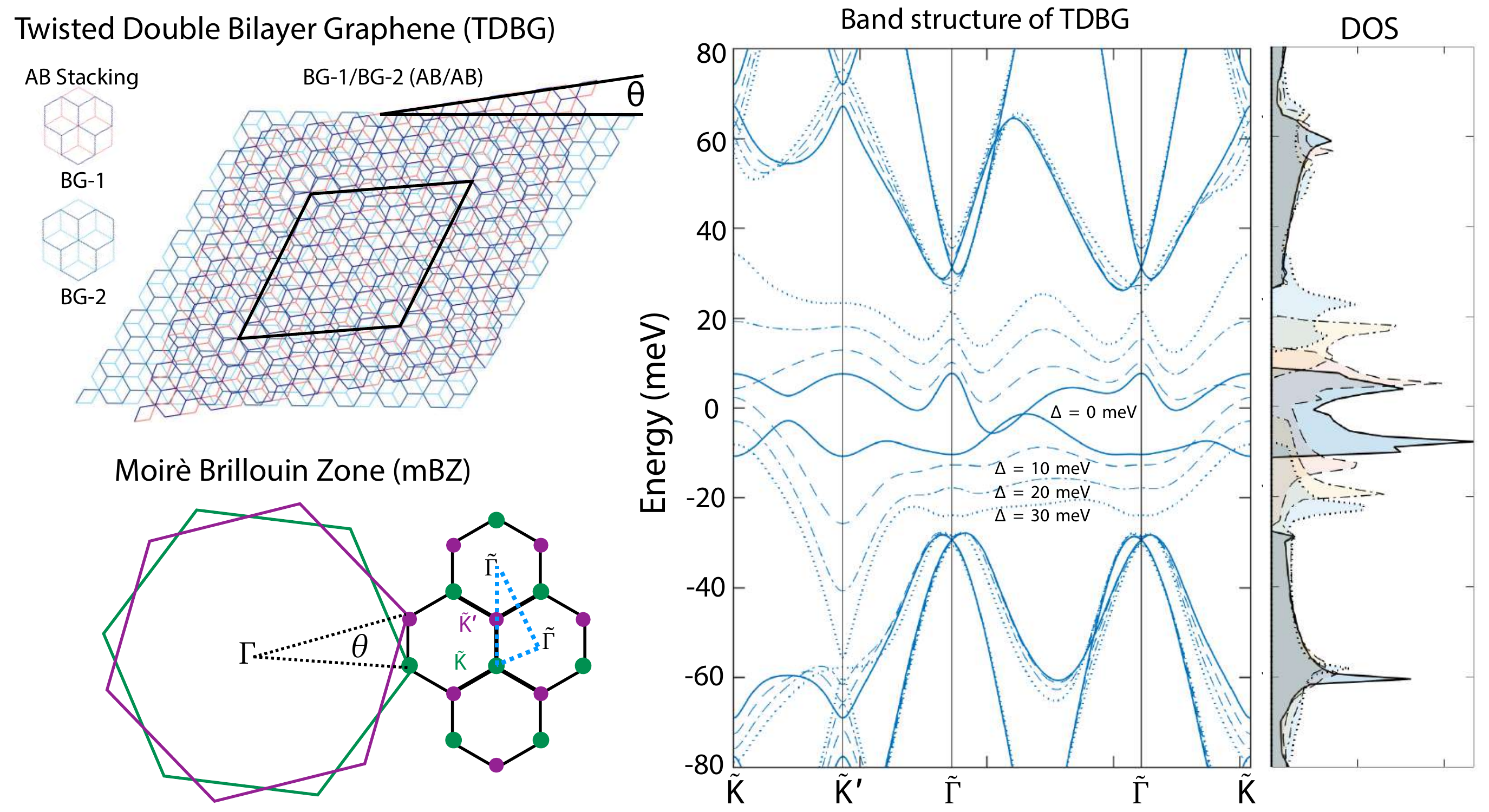}
	\end{center}
	\caption{
			\textbf{Band structure of TDBG.} The twisted double bilayer graphene (TDBG) is the manifestation of two bilayer graphene (BG, i.e., BG-1 and BG-2) with AB stacking on top of each other like BG-1/BG-2 (AB/AB) and rotated along the vertical axis (perpendicular to the plane) with $\theta \neq 0$. The primitive cell of moir\'e cell is represented with solid lines of a hexagonal unit cell. The resulting moir\'e Brillouin zone (mBZ) is formed from the relative displacement of K-points of each bilayer graphene unrotated Brillouin zones. The high symmetry points of mBZ are represented with a tilde to differentiate from the unrotated BZ. We calculated the Band structure along the high symmetry points applied interlayer potential difference ($\Delta$) at a twist angle $\theta = 1.18^{\circ}$. The solid lines are for $\Delta = 0$ meV, dashed lines for $\Delta = 10$ meV, dashed-dot lines for $\Delta = 20$ meV and finally, the dotted lines for $\Delta = 30$ meV. The increase in the $\Delta$ separates the low energy bands near to the charge neutrality point (CNP) opening a primary gap ($\delta_p$) and pushes these bands towards the higher energy bands into the moir\'e gaps between the low energy bands and higher energy bands (secondary gaps, $\delta_s$); eventually these moir\'e gaps get close for large values of $\Delta$. We also represented the density of states (DOS) obtained for different $\Delta$ values, and as represented in the band structures, the different lines in the DOS represent each $\Delta$, as mentioned earlier. For the higher values of $\Delta$, the moir\'e gaps get closed, with a primary gap opening. 
	}\label{supfig1}
\end{figure*}

We can identify the interlayer tunneling with the first harmonic expansion coefficient of the interlayer coupling such that $t_1 = 3\omega$ \cite{jung2014ab, jung-lda-bilayer}, and for simplicity we use the same AB stacking tunneling within each Bernal BG and the twisted interfaces. In the small-angle approximation, the interlayer coupling Hamiltonian is given by 

\begin{equation}
T(\boldsymbol{r}) = \sum_{j=0, \pm} e^{-i \boldsymbol{Q}_j \cdot \boldsymbol{r}} T^j_{l, l'}.
\end{equation}
where the three $\boldsymbol{Q}_j $ vectors $\boldsymbol{Q}_0 = K\theta(0,-1)$ and $\boldsymbol{Q}_{\pm} = K\theta(\pm \sqrt{3}/2,1/2)$ are proportional to twist angle $\theta$ and $K = 4\pi/3a$ is the Brillouin zone corner length of graphene, whose lattice constant is $a = 2.46~\angstrom$, and here the indices $l , l^{\prime}$ label the sublattices of neighboring twisted surface layers. The interlayer coupling matrices between the two rotated adjacent layers are given by 

\begin{equation}
T^0 = \left(\begin{array}{cc}
\omega' & \omega\\
\omega & \omega' \\
\end{array}\right), \hspace{0.3 cm} T^{\pm} = \left(\begin{array}{cc}
\omega' & \omega e^{\mp i 2\pi /3}\\
\omega e^{\pm i 2\pi /3} & \omega' \\
\end{array}\right),
\end{equation}

using a form that distinguishes interlayer tunneling matrix elements $\omega = \omega_{BA^{\prime}}$ and $\omega^{\prime} = \omega_{AA^{\prime}}$ for different and same sublattice sites between the layers. The convention taken here for the $T^j$ matrices \cite{jung2014ab} assumes an initial AA stacking configuration $\tau = (0, 0)$ and differs by a phase factor with respect to the initial AB stacking $\tau = (0, a/\sqrt3)$~\cite{gapdirac-srivani,necolas-prb}. The greater interlayer separation $c$ compared to the carbon-carbon distances $a_{CC}$ lead to slowly varying interlayer tunneling function $T(\boldsymbol{r})$ and the moir\'e patterns can often be accurately described within a first-harmonic approximation \cite{bistritzer2011moire, jung2014ab}. Furthermore, the effects of atomic relaxation in the moir\'e patterns lead to corrugations that have non-negligible effects in the details of the electronic structure for both intralayer potentials and interlayer coupling \cite{jung-natcum}. In the case of $\omega' \neq \omega$, we proposed a single parameter relation through $\omega' = C_1 \omega^2 + C_2 \omega + C_3$, where $C_1 = -0.5506$, $C_2 = 1.036$, $C_3 = -0.02245$ as discussed in Ref. \cite{tbbg-raju} (in its Supplementary information). Our calculations have used a configuration space with variable cutoff in momentum space of a radius of up to $6G_1 = 24\pi \theta/(\sqrt{3}a)$ using Hamiltonian matrices with sizes as large as $676 \cross 676$ such that $\theta \gtrsim \omega/(12\pi|t_0|)$ to obtain converged results in the limit of small $\theta$ and large $\omega$.

\begin{figure*}[hbt]
	\begin{center}
		\includegraphics[width=16 cm]{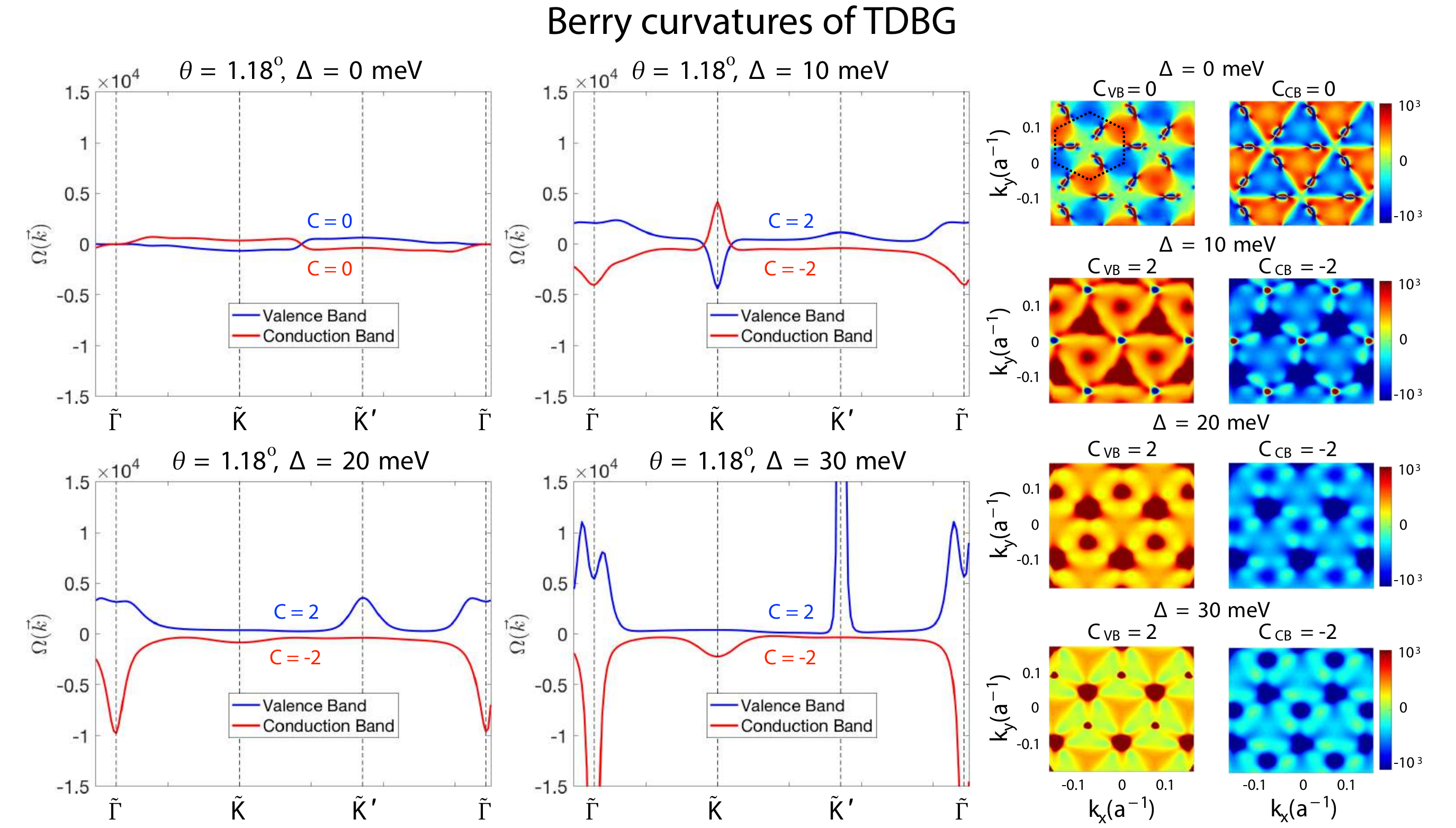}
	\end{center}
	\caption{
			\textbf{Berry curvatures of TDBG.} We calculate the Berry curvatures ($\Omega$) for TDBG at a twist angle $\theta = 1.18 ^{\circ}$ for the different applied interlayer potential ($\Delta$). On the left side, we represent the Berry curvature lines along with the high symmetry points for the different values of $\Delta$. The resultant Chern values for the low energy bands are represented for each $\Delta$ value with color representation similar to the valence and conduction band. The Berry curvature did not show singularities behavior for the $\Delta = 0$ meV, with resulting $C_\text{VB/CB} = 0$. However, for the non zero $\Delta$, the Berry curvatures have singularities at high symmetry points at $\tilde{K}$ and $\tilde{\Gamma}$ for the conduction band, and at $\tilde{K}^{\prime}$ and $\tilde{\Gamma}$ for valence bands, the weight of the singularities varied with applied $\Delta$ values. On the right side, the surface plot for the entire mBZ is represented for each case of $\Delta$ for the valence and conduction bands. The colorbars on the right represents $\Omega$ in arbitrary units.
	}\label{supfig2}
\end{figure*}

The possibility of band gap opening at charge neutrality point (CNP), primary gap ($\delta_p$), through an electric field together with the presence of moir\'e gaps (secondary gaps, $\delta_s$) with the higher-energy bands leads to well-defined valley Chern numbers. The valley Chern numbers were calculated through 
\begin{equation}
C_{\upsilon} =  \int_\textrm{mBZ} d^2 \vec{k} ~ \Omega_{q} (\vec{k})/{2 \pi},
\end{equation}
by integrating the moir\'e Brillouin zone for each valley the Berry curvature for the $q$-th band through~\cite{rmp-berry}
\begin{equation}
\Omega_{q} (\vec{k}) = -2 \sum_{q' \neq q} \textrm{Im} \Bigg[ \frac{\langle u_{q} \vert \frac{\partial H}{\partial k_x} \vert u_{q^{\prime}} \rangle \langle u_{q^{\prime}} \vert \frac{\partial H}{\partial k_y} \vert u_{q} \rangle}{(E_{q^{\prime}} - E_{q})^2} \Bigg],
\label{Bcurv}
\end{equation}

where for every k point we take sums through all the neighboring $q^{\prime}$ bands, the $\vert u_{q} \rangle$ are the moir\'e superlattice Bloch states, and $E_{q}$ are the eigenvalues.

\begin{figure}[hbt]
	\includegraphics[width=9cm]{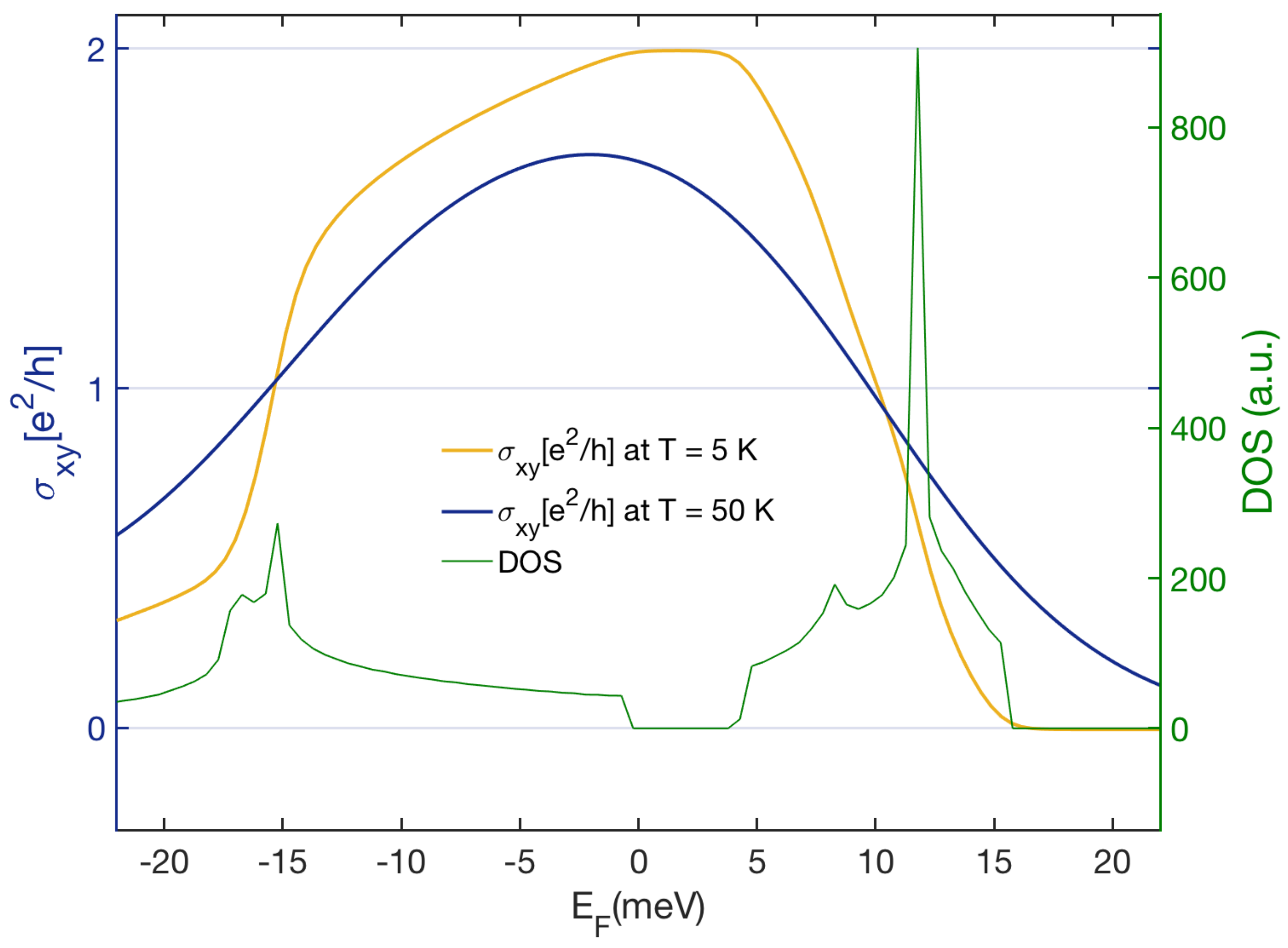}
	\caption{ \label{fig:sigmaxytheory}  \textbf{Hall conductivity at the CNP.} The variation of Hall conductivity ($\sigma_{xy}$) (on left axis) for the $1.18^{\degree}$ TDBG device as a function of Fermi energy ($E_\text{F}$) at $T=5$~K (orange curve) and $T=50$~K (blue curve). The corresponding variation of Density of states (on right axis) across the charge neutrality is overlaid (green curve). The two peaks in the DOS, towards the positive and negative sides of $E_\text{F}$, correspond to the two flat bands separated by a band gap close to $E_\text{F}=0$. Here, the interlayer potential difference chosen to open up a gap at the CNP between the flat bands is $\Delta=15$~meV. }
\end{figure}

The Hall conductivity that results due to a nonzero Berry curvature ($\Omega$) at a particular valley of graphene (K or K$^\prime$) is given by~\cite{sigmaxy_formula}

\begin{equation}
\sigma_{xy} (E_\text{F})=  \frac{1}{2\pi}\frac{e^{2}}{h}\sum_{q}\int d^2 \vec{k} ~ \Omega_{q} (\vec{k}) f(\epsilon_{q}(\vec{k}),E_\text{F}),
\end{equation}
where $q$ indicates the bank index and $f(\epsilon_{q}(\vec{k}))$ denotes the Fermi occupation function. At low temperature (shown in orange curve for $T=$~5~K in Fig.~\ref{fig:sigmaxytheory}), it saturates to $\frac{2e^2}{h}$ in the band gap between the flat bands. Away from the gap, it starts decreasing since the Berry curvature in the low energy conduction band and valence band have opposite signs, as seen from the line plots in Fig.~\ref{supfig2}. This decrease of $\sigma_{xy}$ away from the CNP gap in TDBG is asymmetric in nature, which is in contrary to that obtained for bilayer graphene~\cite{shimazaki_generation_2015-1_s,koshino_bilayer}. The Hall conductivity is also shown for an elevated temperature of $T=$~50~K (blue curve) where we have observed bulk valley transport at the CNP gap. The decrease in $\sigma_{xy}$ from its low temperature value of $\frac{2e^2}{h}$ at the gap is due to the thermal excitation of valence band electrons to the conduction band. The valley Hall conductivity, $\sigma_{xy}^{\text{VH}}$, is obtained by adding the contribution to the Hall conductivity from the individual valleys at K and K$^\prime$, and is given by $\sigma_{xy}^{\text{VH}}=2\sigma_{xy}$~\cite{shimazaki_generation_2015-1_s}.   

%\clearpage
%\section{Details of device fabrication}

%Our dual gated devices are made of hBN/TDBG/hBN stacks on SiO$_2$(around 280~nm)/Si$^{++}$ substrate. To make the stacks we exfoliated graphene and chose suitable bilayer graphene flakes based on optical contrast and then confirmed the layer number by Raman spectroscopy. 
%The suitable hBN flakes were chosen based on color, and we confirmed the thickness by AFM after the stack is completed. Bilayer graphene flakes were sliced into two halves using a tapered optical fiber scalpel, a method reported in~\cite{sangani_facile_2020}. Subsequently the flakes were assembled using the standard poly(propylene) carbonate (PPC) based dry transfer method~\cite{wang_one-dimensional_2013-1}. The twist angle was introduced by rotating the bottom stage during the pick up of the second half of the graphene.
%Subsequently we defined the geometry of the devices by e-beam lithography followed by  CHF$_3$ + O$_2$ plasma etching. One-dimensional edge contacts to the graphene were made by etching the stack and depositing Cr/Pd/Au. The top gate was made by depositing Cr/Au. 

%We fabricated and measured nonlocal transport in multiple devices. All the data reported in the main manuscript are measured using Device 1 with twist angle 1.18$\degree$. We present data from another device with twist angle 1.24$\degree$ in Sec.~\ref{sec:dev2} of this Supplementary text.

%\clearpage 

\section{Nonlocal transport measurement}\label{sec:meas}
\begin{figure}[hbt]
	\includegraphics[width=15cm]{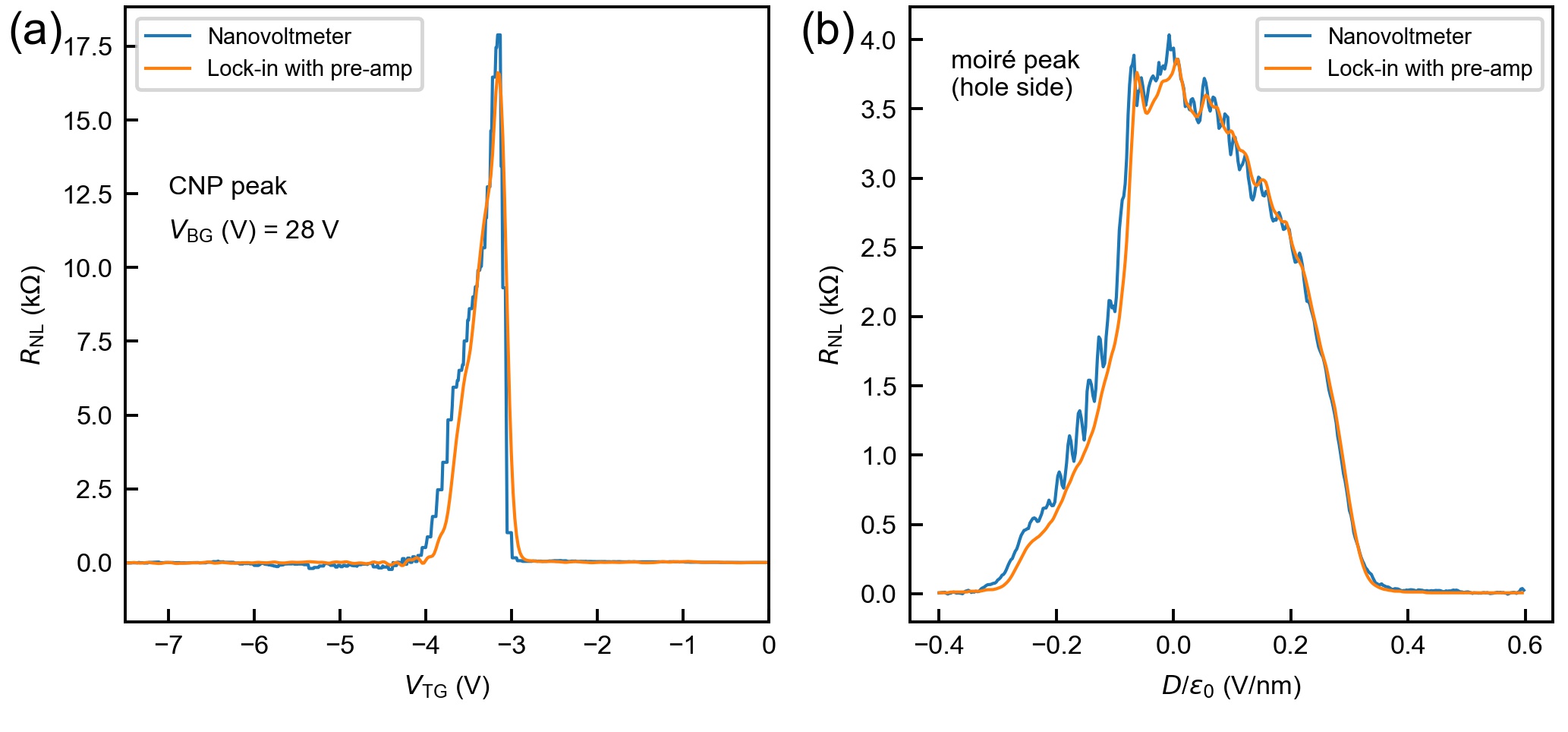}
	\caption{ \label{fig:meas_sceme}  \textbf{Nonlocal resistance measurement using different setups.} (a) Nonlocal resistance as a function of $V_\text{TG}$ at CNP peak for two different measurement schemes -- dc measurement using Keithley 2182 nanovoltmeter along with  Keithley 6221 current source and ac measurement using lock-in technique with a preamplifier. (b) Nonlocal resistance as a function of the electric field for $n=-n_\text{S}$ using the two methods.}
\end{figure}

The transport measurement reported in the main text and all other sections in the supplementary was measured using a low frequency ($\sim$ 17 Hz) lock-in technique by sending a current $\sim$ 10 nA and measuring the voltage after amplifying using SR560 preamplifier or preamplifier model 1021 by DL instruments, Ithaca.   Since the nonlocal resistance is appreciable when the system is gapped, spurious signal can be measured for these high resistivity states~\cite{sui_gate-tunable_2015-1_s, shimazaki_generation_2015-1_s}. To verify the nonlocal resistance we measured is not a measurement artifact, we employ Keithley 2182 nanovoltmeter to measure dc voltage while sending current using Keithley 6221 current source. The nanovoltmeter has input impedance > 10 G$\Omega$ while the SR560 or DL 1021 preamplifier has an input impedance of 100 M$\Omega$.  In Fig.~\ref{fig:meas_sceme}(a) we have plotted the nonlocal resistance  as a function of top gate voltage $V_\text{TG}$ around the CNP at $V_\text{BG} = 28$ V measured both by  the lock-in method aided by the preamplifier and the dc measurement using nanovoltmeter. For measurement using the nanovoltmeter, we measured resistance both in forward and reverse directions and took the average resistance to nullify any dc voltage drop due to thermo-electric effect at various junctions of the current path inside the cryostat. In Fig.~\ref{fig:meas_sceme}(b) we plotted the nonlocal resistance of the \moire peak at the hole side as a function of the perpendicular electric field using the two schemes. As seen from both Figs.~\ref{fig:meas_sceme}(a) and \ref{fig:meas_sceme}(b),  the resistance values are independent of the measurement schemes. All the measurements subsequently were done using lock-in with the preamplifier. 

%\clearpage
\section{Angle homogeneity for nonlocal detection of valley current}\label{sec:angle}

\begin{figure}[h]
	\includegraphics[width=8.5cm]{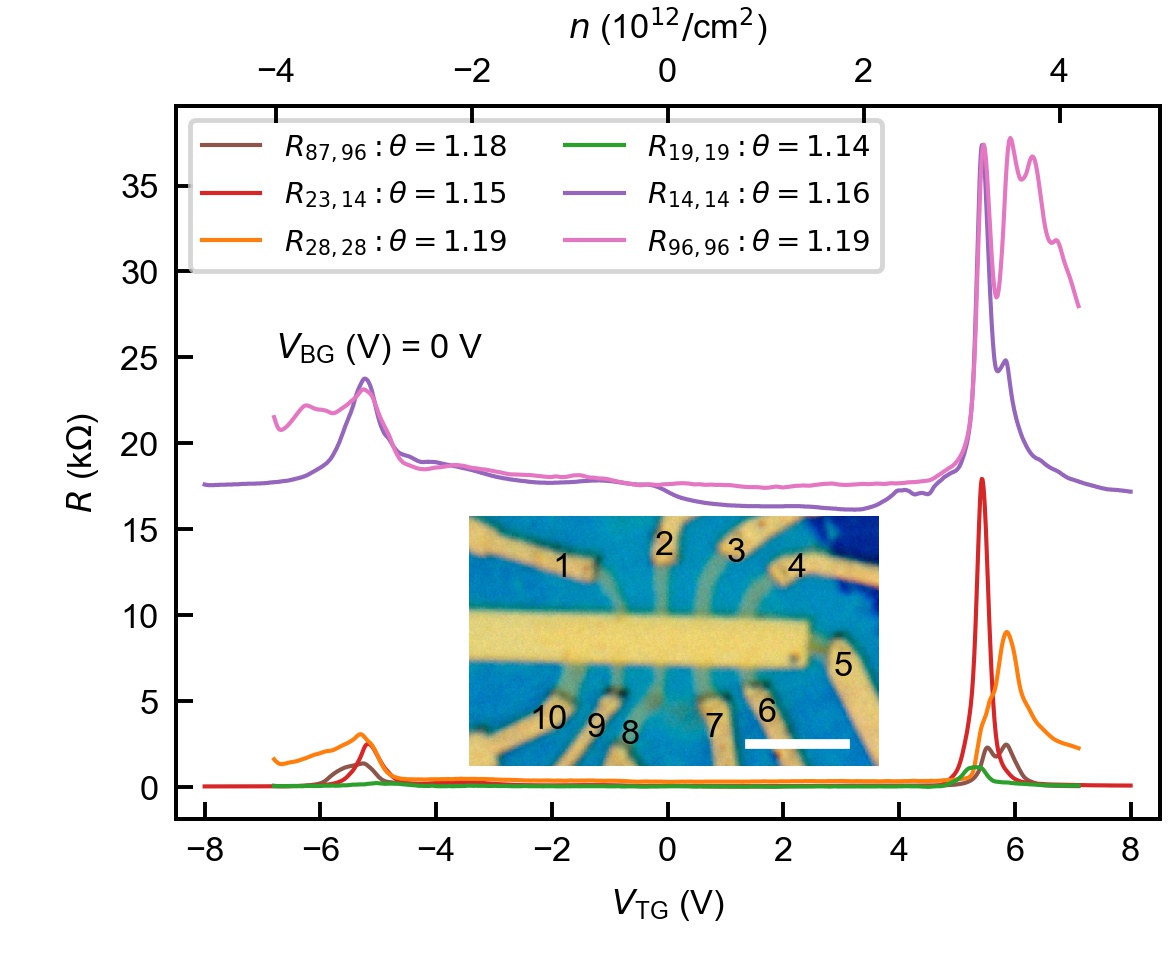}
	\caption{ \label{fig:angle_inhomo}  \textbf{Estimating angle inhomogeneity.} Variation of two-probe and four-probe local resistance using different terminals of the device as a function of $V_\text{TG}$ at $V_\text{BG} = 0$ V. In $R_{ij,kl}$, the indices $i,j$ denote the voltage probes and $k,l$ denote the current probes. The difference between the positions of two \moire peaks is used to estimate the local twist angle. Inset: An optical microscope image of the device with the terminals numbered. The scale bar corresponds to 5~\textmu m.}
\end{figure}
The twist angle homogeneity is an important prerequisite for measuring nonlocal valley transport in twisted graphene devices. This is because the generation of valley current in the injection probes happens when the Fermi energy lies in a Berry curvature hotspot, i.e., the Fermi energy lies near a gap. This requires the charge density to be tuned by the gates to specific values: $n =  0$~(CNP gap) or $n = \pm n_\text{S}$~(\moire gap), where $n_\text{S}=8\theta^{2}/(\sqrt{3}a^{2})$ with  $a=0.246$~nm being the lattice constant of graphene. Now for detecting the valley current, the detection probes have the same requirement. Since $n_\text{S}$ depends on $\theta$, the local twist angle should be the same in both the pairs of injection and detection probes as well as the valley current path.

To estimate the local twist angle near various probes we measure two-probe and four-probe resistance using different combinations of current and voltage probes. In Fig.~\ref{fig:angle_inhomo} we present  such different plots of  local resistance as a function of $V_\text{TG}$ for $V_\text{BG} = 0$~V. All the curves have two \moire peaks corresponding to $n = \pm n_\text{S}$. The difference in the positions of the two peaks on the $n$-axis corresponds to $2 n_\text{S}$, which is used to estimate the local twist angle. We find that the twist angle to vary between 1.14$\degree$ to  1.19$\degree$, establishing that the device has good angle homogeneity. 

%\clearpage
\section{Reciprocity of the nonlocal resistance measurement}\label{sec:reciprocity} 
\begin{figure}[hbt]
	\includegraphics[width=8cm]{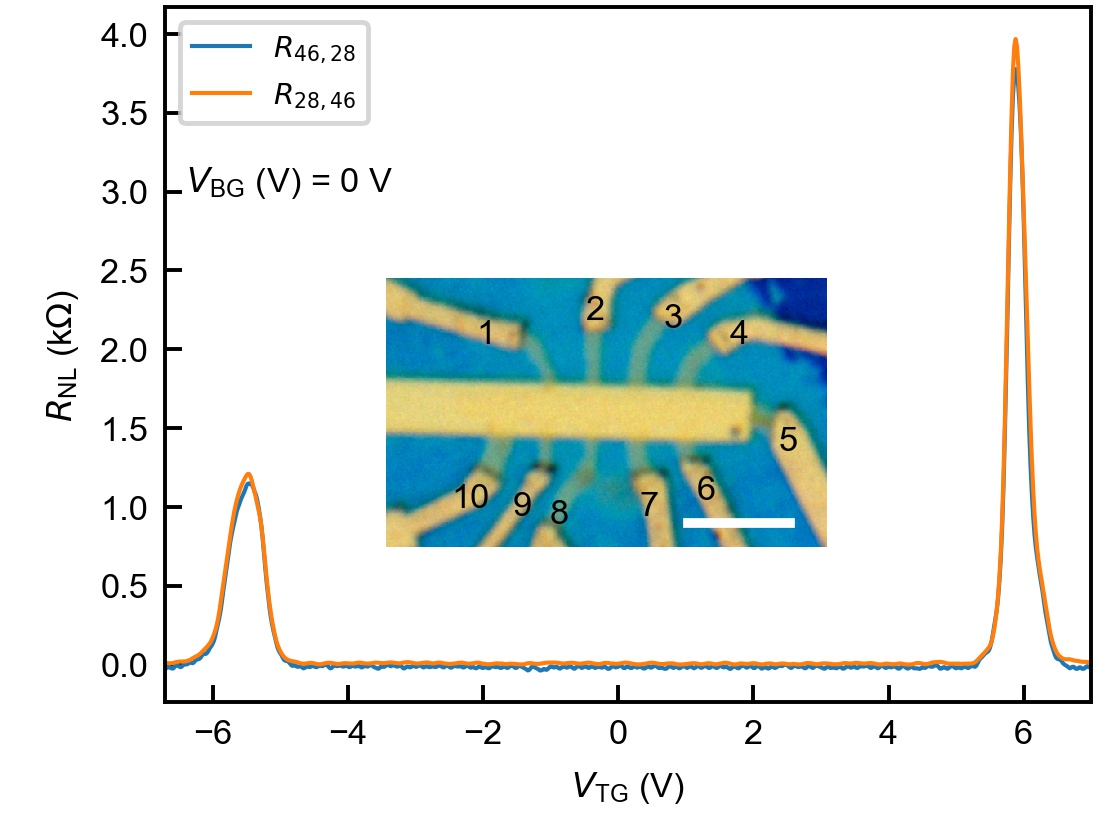}
	\caption{ \label{fig:reciprocity}  \textbf{Verification of reciprocity of the nonlocal resistance measurement.} Nonlocal resistance as a function of $V_\text{TG}$ with $V_\text{BG} = 0 $~V for two reciprocal combinations of injection and detection probes. In $R_{ij,kl}$, the indices $i,j$ denote the voltage probes (detection terminals) and $k,l$ denote the current probes (injection terminals). The inset shows the image of the device with the scale bar corresponding to 5 \textmu m. }
\end{figure}

For the Onsager reciprocal relations to be valid, the nonlocal resistance we measure should be the same if one swaps the injection and the detection terminals~\cite{shimazaki_generation_2015-1_s}. We verify this in Fig.~\ref{fig:reciprocity} where we present nonlocal resistance  as a function of $V_\text{TG}$ with $V_\text{BG} = 0 $~V for two reciprocal combinations of injection and detection probes. We find that the nonlocal resistance does not change if we swap the terminals.

%\clearpage

\section{Additional data on scaling}\label{sec:scaling}

\begin{figure}[hbt]
	\includegraphics[width=16cm]{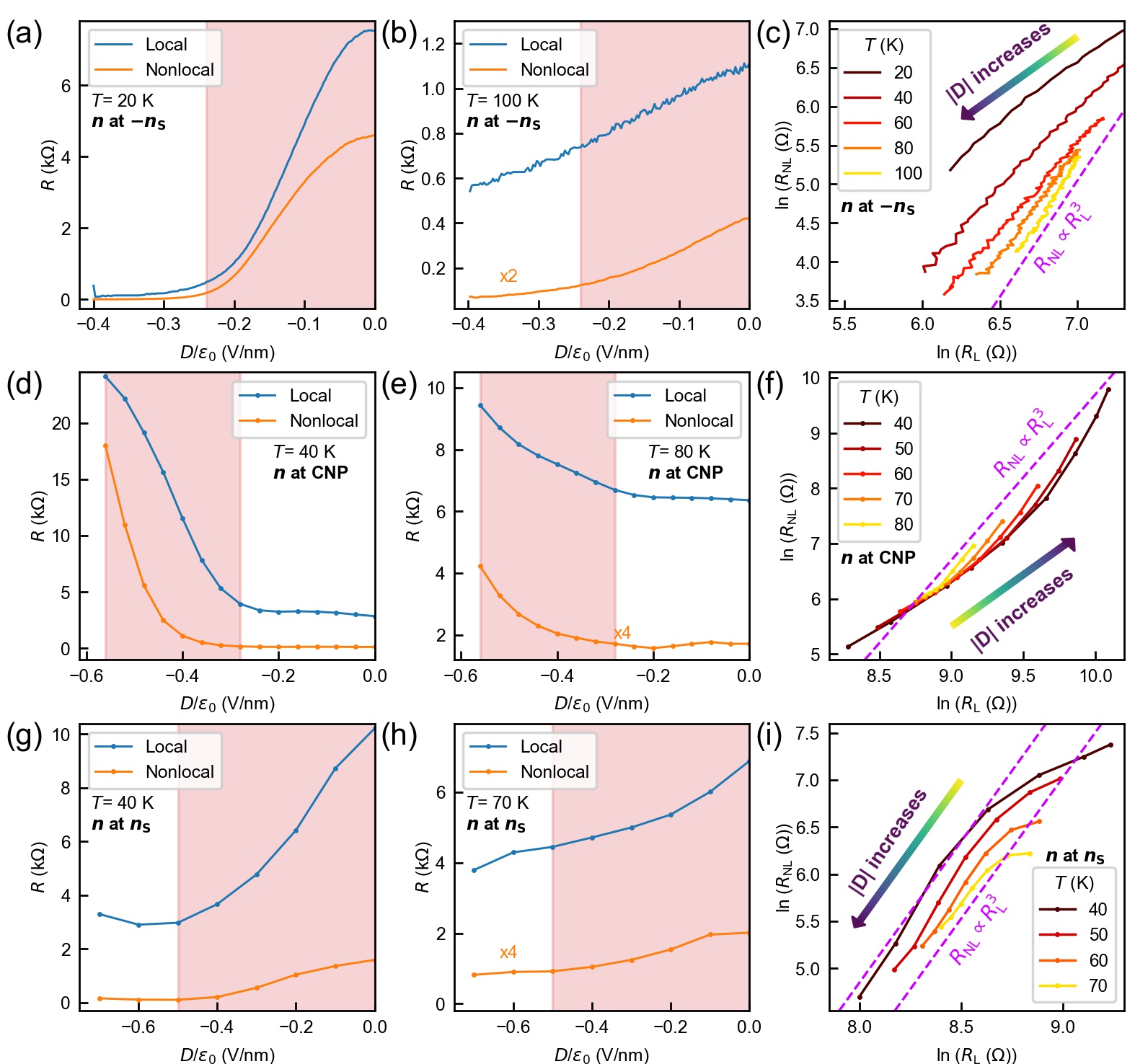}
	\caption{ \label{fig:Dscaling}  \textbf{Scaling of nonlocal resistance with local resistance at fixed temperatures by taking electric field as parameter.} (a-b,d-e,g-h) The variation of nonlocal (orange) and local (blue) resistance as a function of the perpendicular electric field for doping at $-n_\text{S}$~(a-b), $n=0.04\cross10^{12}~\text{cm}^{-2}$~(CNP)~(d-e), and $n_\text{S}$~(g-h). The variation is plotted for fixed temperatures that are indicated in each plot. The nonlocal resistances in (b,e,h) are plotted after multiplying by constant factors that are indicated in orange. The pink background indicates the electric field range used to show scaling in (c,f,i). (c,f,i) The scaling of nonlocal resistance with local resistance with the electric field as the parameter. The scaling is shown for doping at $-n_\text{S}$~(c), $n=0.04\cross10^{12}~\text{cm}^{-2}$~(CNP)~(f), and $n_\text{S}$~(i) for various fixed temperatures. The dashed violet line is a guide to the eye for cubic scaling. The arrows indicate the direction of the rise in the magnitude of the electric field.}
\end{figure}
A tell-tale signature of bulk valley current is cubic scaling between the nonlocal and local resistances. In Fig.~3e and Fig.~3f of the main text, we had demonstrated the cubic scaling by taking temperature as a parameter for some fixed electric fields. In Fig.~\ref{fig:Dscaling}, we show the cubic scaling at different fixed temperatures for the CNP (Fig.~\ref{fig:Dscaling}(f)) and \moire gaps (Fig.~\ref{fig:Dscaling}(c) for $n=-n_\text{S}$ and Fig.~\ref{fig:Dscaling}(i) for $n=n_\text{S}$) by taking the electric field ($D$) as a parameter. For $n=-n_\text{S}$ in Fig.~\ref{fig:Dscaling}(c), we see cubic scaling towards the high temperature end. The case for $n=n_\text{S}$ in Fig.~\ref{fig:Dscaling}(i) shows cubic scaling in the high electric field end, while the scaling deviates from cubic and saturates towards the low electric field end where the band gap at $n=n_\text{S}$ is higher~\cite{adak_tunable_2020_prb_s}. This saturation at high band gap regime is attributed to large valley Hall angle physics \cite{polini} and is also seen in earlier studies at the CNP gap of bilayer graphene \cite{shimazaki_generation_2015-1_s}. In Fig.~\ref{fig:Dscaling}(f) we find that the CNP shows cubic scaling at high temperatures. As the temperature is lowered (brown and black curve in Fig.~\ref{fig:Dscaling}(f)), the scaling deviates from cubic to higher exponents in the high electric field end. This is similar to that in Fig.~3(f) of the main text, where high $D$ and low $T$ shows the same departure.

%\clearpage

\section{Data from the second device}\label{sec:dev2}
\begin{figure}[hbt]
	\includegraphics[width=15cm]{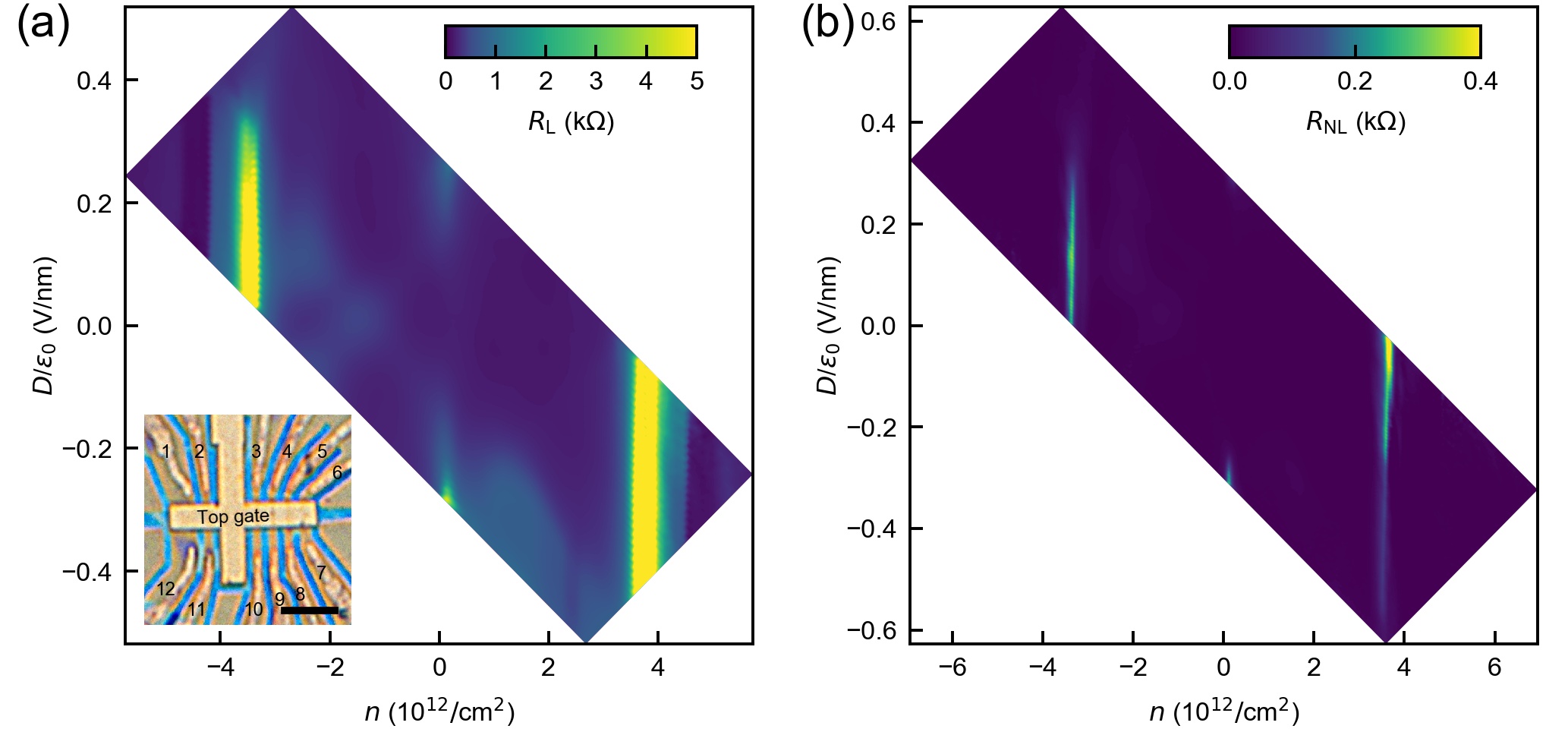}
	\caption{ \label{fig:device2}  \textbf{Nonlocal resistance measured in device two with twist angle $1.24\degree$ at 1.5 K.} (a)~Color scale plot of local resistance as a function of charge density and electric displacement field. Inset: Optical microscope image of the device with electrodes labeled by numbers. Scale bar corresponds to 4 \textmu m. (b)~Color scale plot of nonlocal resistance as a function of charge density and electric displacement field. For measuring the nonlocal resistance, the current was sent through the terminals 9 and 4, and voltage was measured across the terminals 11 and 2.}
\end{figure}

In Fig~\ref{fig:device2}, we present local and nonlocal resistance from the Device 2 which has a twist angle of $1.24\degree$. 
%\textcolor{red}{To further verify that the nonlocal resistance we measure originate from bulk valley transport we employ a different geometry in this device. As seen from the device image of the insets in  Fig~\ref{fig:device2} the device has an elongated nose. The nonlocal resistance is measured with probe combination such that any current via edge of the sample has to traverse an additional path of 6 \textmu m, by which the edge current should have decayed. The significant nonlocal resistance measured then must be originating from the bulk current. \textbf{Comment: this has message condradictory to that presented in the main text. It supports against edge transport, while in the main text we say that at low temperature there might be contribution from the edge current. Should we include this part? }}

%\clearpage
\section{Decay of Nonlocal resistance with length}
\begin{figure}[hbt]
	\includegraphics[width=15cm]{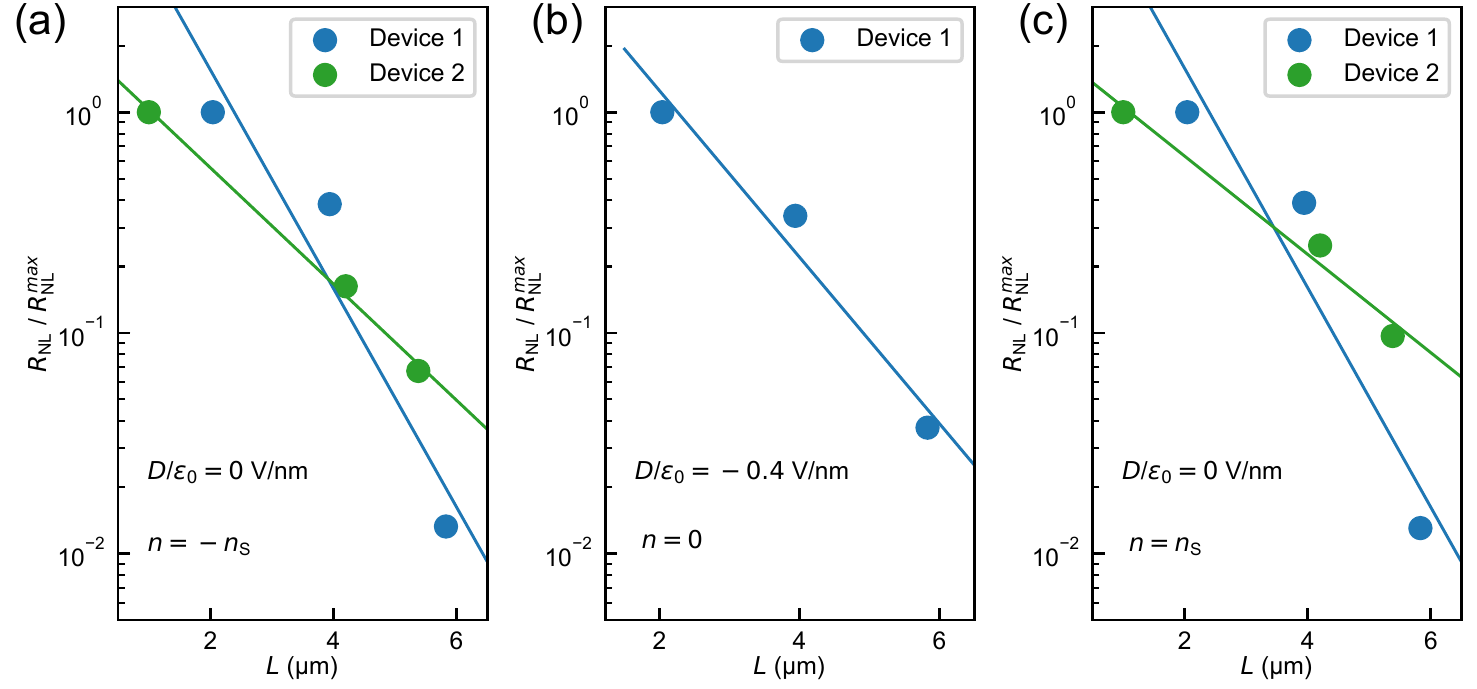}
	\caption{ \label{fig:diffusion_length}  \textbf{Decay of nonlocal resistance over length.} (a-c) The variation of the nonlocal resistance as a function of length for $n=-n_\text{S}$ (a), $n = 0$ (b) and $n=n_\text{S}$ (c) for Device~1 and Device~2. Here, $R_\text{NL}$ has been normalized with respect to the nonlocal resistance measured in the nearest probe,~$R_\text{NL}^\textit{max}$, for both the devices.}
\end{figure}

The dependence of the nonlocal resistance on the length of the current path is governed by the valley diffusion length $l_\text{v}$, as seen from equation~(\ref{eqn:RnlvsRl}), 
\begin{equation}\label{eqn:RnlvsRl}
R_\text{NL}=\frac{1}{2} \left( \frac{\sigma_{xy}^{\text{VH}}}{\sigma_{xx}}\right)^2 \frac{W}{\sigma_{xx}l_\text{v}} \exp(-\frac{L}{l_\text{v}}).
\end{equation}
Here, $\sigma_{xy}^{\text{VH}}$ is the valley Hall conductivity. 
$L$ and $W$ represent the length and width of the Hall bar channel, respectively. To extract $l_\text{v}$ we plot the decay of the nonlocal signal along the sample length at the \moire gaps and the CNP in Fig.~\ref{fig:diffusion_length}.  We fit the exponential decay using equation~(\ref{eqn:RnlvsRl}) and extract out $l_\text{v}$ for both the devices. The data for Device~1 is plotted at 35~K to negate low temperature effects.  For \moire gaps, we find $l_\text{v}$ to be  0.9 \textmu m and 1.8 \textmu m for Device~1 and Device~2 respectively. At the CNP,  $l_\text{v}$ is 1.15 \textmu m  for Device~1. These values are similar to those obtained in bilayer graphene~\cite{shimazaki_generation_2015-1_s}. 

%\clearpage
%\section{Temperature dependence of valley Hall conductivity}

\section{Effect of band flatness on Berry curvature hotspot}\label{sec:berry}
\begin{figure}[hbt]
	\includegraphics[width=15cm]{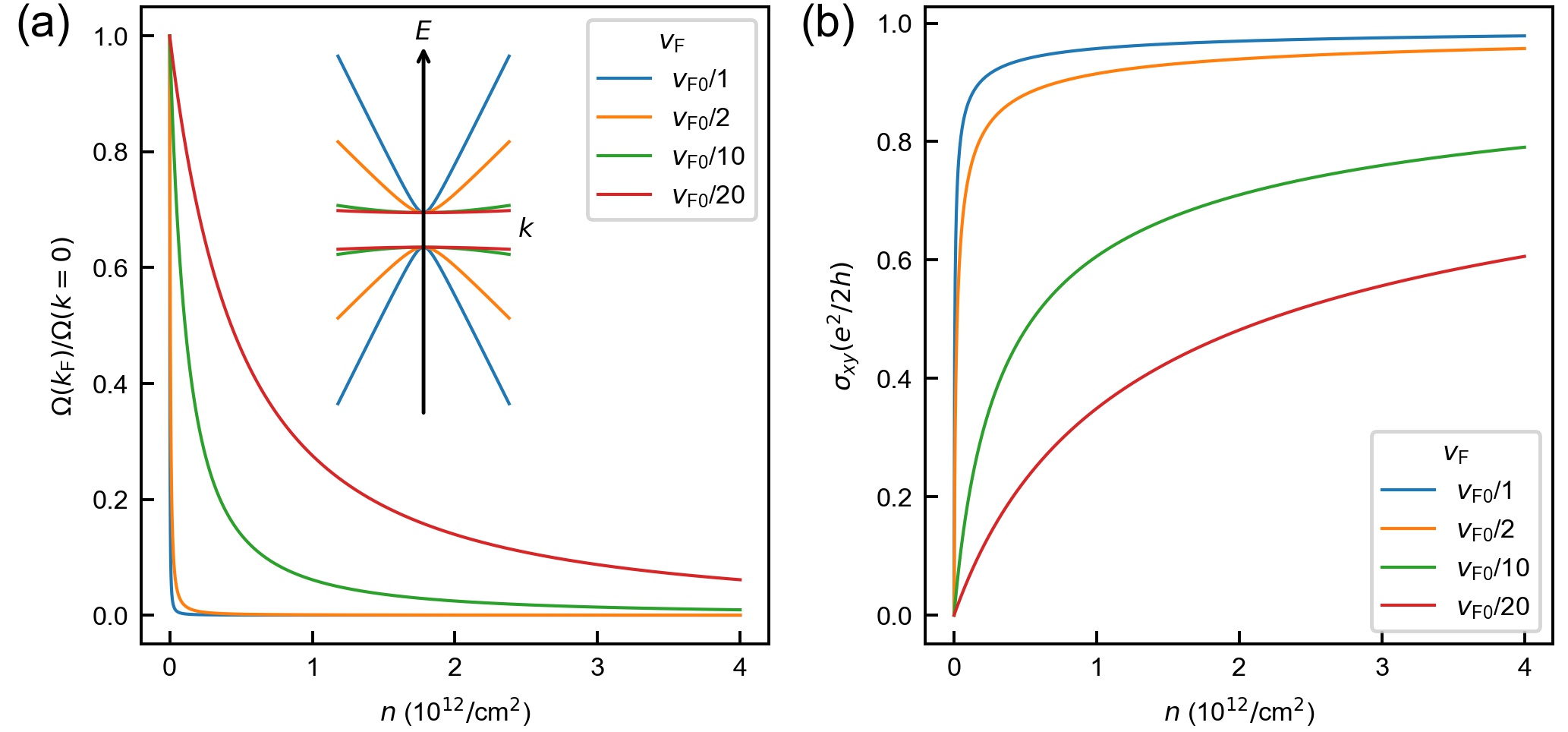}
	\caption{ \label{fig:berry_curvature}  \textbf{Effect of the flat band on Berry curvature hotspot.} (a) The variation of the Berry curvature at the Fermi wave vector $k_\text{F}$ as a function of charge density in the gapped monolayer graphene band for different Fermi velocities. The band gap used for the calculation is 10 meV. The inset shows the band dispersion. (b) The dependence of valley Hall conductivity as a function of charge density for different Fermi velocities. The band gap is 10 meV.}
\end{figure}

To understand the effect of a flat band on the Berry curvature hotspot, we consider a simple band structure of gapped monolayer graphene.
We choose the monolayer graphene Hamiltonian since, for a flat band, in any twisted system, there is no reason in general for the linear term in momentum to be zero.
The Hamiltonian for the gapped monolayer graphene is given by, $H=\hbar v_\text{F} \bm{\sigma}.\bm{k}+\Delta_\text{g} \sigma_3$.
%\begin{equation}
%H = \hbar v_\text{F}
%\begin{bmatrix}
%\delta & k_x-ik_y\\
%k_x+ik_y & -\delta 
%\end{bmatrix}
%\end{equation}
Here $2\Delta_\text{g}$ is the band gap between two energy bands given by $E_{\pm} = \pm  \sqrt{ (\hbar v_\text{F} k)^2 +\Delta_\text{g}^2}$. 
We incorporate the band flatness by renormalizing the term $v_\text{F}$ in the Hamiltonian, whereas in the case of monolayer graphene $v_\text{F} = 10^6$ m/s $= v_\text{F0}$. 
The Berry curvature is given by $\Omega (k) = (\hbar v_\text{F})^2 \Delta_\text{g}/[2 ((\hbar v_\text{F} k)^2+\Delta_\text{g}^2)^{3/2}]$. 
With the charge density $n$ given by $n = k_\text{F}^2/\pi$ and $\delta=\Delta_\text{g}/\hbar v_\text{F}$, the Berry curvature at the Fermi wave vector $k_\text{F}$ is given by $\Omega (k_\text{F}) = \delta/[2 ( n\pi + \delta^2)^{3/2}]$. 
%So the extent of the Berry curvature hotspot is governed by $\delta$ in the denominator.
%This means that for the same value of gap $\Delta$, if $v_\text{F}$ decreases $\delta$ increases and hence the hotspot extends more.
In Fig.~\ref{fig:berry_curvature}(a) we plot the Berry curvature as a function of charge density for monolayer graphene with a gap of $2\Delta_\text{g} = 10$~meV and for different values of the renormalized velocity, $v_\text{F}$. 
As $v_\text{F}$ is smaller, i.e., the band is flatter, we find that the Berry curvature is more delocalized away from the gap. 
In Fig.~\ref{fig:berry_curvature}(b) we plot the contribution of the conduction band to the valley Hall conductivity given by, $\sigma_{xy} = (e^2/2h)\int \Omega(k) k dk d\theta =   (e^2/2h)(1-\delta/\sqrt{n\pi+\delta^2})$. We again find that in the case of flat band the Hall conductivity saturates to its asymptotic value much slower.
The Berry curvature delocalization and slow saturation of Hall conductivity in flat band systems, as represented in Fig.~\ref{fig:berry_curvature}, can be reproduced even by starting with a gapped bilayer graphene Hamiltonian.
%The contribution of the conduction band to the valley Hall conductivity is given by, $\sigma_{xy} = (e^2/2h)\int \Omega_C(k) k dk d\theta =   (e^2/2h)(1-\delta/\sqrt{n\pi + \delta^2})$. 
%In Fig.~\ref{fig:berry_curvature}(b) we plot the Hall conductivity as a function of the charge density $n$.
%We again find that in the case of a flat band the decay of magnitude of the Hall conductivity is much slower.

%\clearpage

\section{Comparison of extent of nonlocal resistance peak with a system without flat bands}\label{sec:compare}

\begin{figure}[hbt]
	\includegraphics[width=8.5cm]{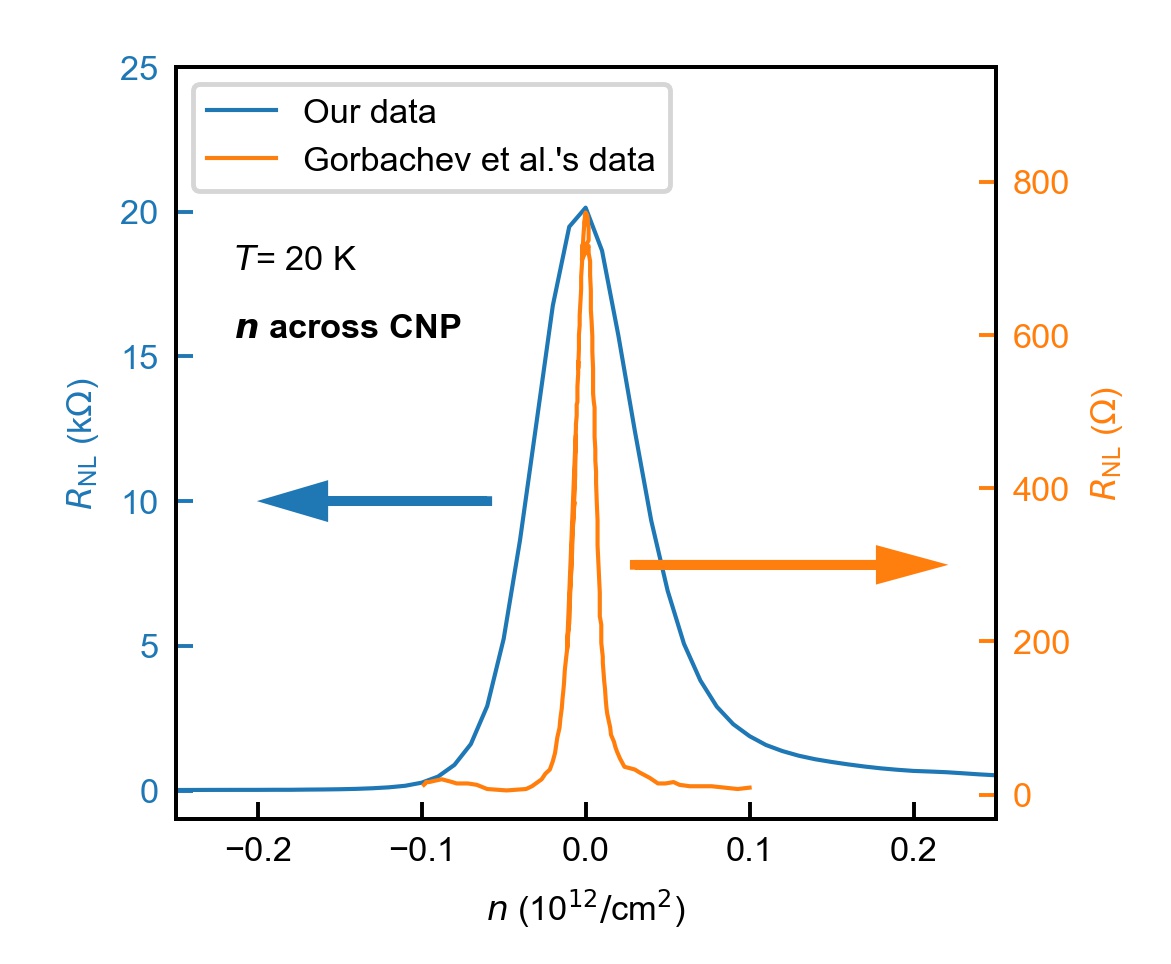}
	\caption{ \label{fig:Rnlvsn} \textbf{Effect of the flat band on the extent of nonlocal resistance peak.} The variation of the nonlocal resistance as a function of charge density varying across the charge neutrality point. The blue curve represents the measured nonlocal resistance at the CNP gap at a perpendicular electric field of $-0.4$~V/nm in our 1.18\textdegree~TDBG device at $T$=~20~K. The orange curve is the nonlocal resistance variation across the charge neutrality point from hBN aligned graphene device, which does not have any flat band. The data is used from Fig.~S3(B) of Gorbachev~et~al.~\cite{gorbachev_detecting_2014_s}, corresponding to $T$=~20~K.}
\end{figure}

Here we compare the extent of the nonlocal signal in the charge density axis from our TDBG device with that of hBN aligned graphene device from Gorbachev~et~al.~\cite{gorbachev_detecting_2014_s}. We compare the charge neutrality peak at $T$=~20~K for both the devices. Upon application of a perpendicular electric field (shown for $D/\epsilon_0=-0.4$~V/nm in Fig.~\ref{fig:Rnlvsn}), a gap opens up at charge neutrality within the flat bands in the TDBG device~\cite{adak_tunable_2020_prb_s}. These bands have Berry curvature hotspots, as shown in Fig.~\ref{supfig2}. For the case of hBN aligned graphene, the superlattice potential results in opening up a gap between the valence band and conduction band at the charge neutrality~\cite{song_topological_2015_s}. This, too, results in Berry curvature hotspots at the band edges near the gap opening. However, unlike the case in small-angle TDBG, the resulting low energy bands are not flat. The broader extent of the nonlocal resistance in our data (the FWHM being $\sim$5 times larger) can be attributed to the spreading of Berry curvature hotspot due to flat bands in the TDBG system.

%\clearpage

%\bibliography{bibs}
\input{bibs.bbl}
\end{document}